\documentclass[twocolumn,apjfonts,tighten]{aastex631}

\usepackage{apjfonts}
\usepackage{graphicx}
\usepackage{wrapfig}
\usepackage{amsmath}
\usepackage{multirow}
\usepackage{booktabs}
\usepackage{float}
\usepackage{gensymb}

\defcitealias{lewis_panchromatic_2015}{L15}

\shorttitle{PHAST. II. The Recent Star Formation History of M31}
\shortauthors{Wainer et al.}

\begin{document}

\title{The Panchromatic Hubble Andromeda Southern Treasury (PHAST). II. \\ 
The Spatially Resolved Recent Star Formation History in M31}

\author[0000-0001-6320-2230]{Tobin M. Wainer}
\affiliation{Department of Astronomy, University of Washington, Box 351580, Seattle, WA 98195, USA}

\author[0000-0002-7502-0597]{Benjamin F. Williams}
\affiliation{Department of Astronomy, University of Washington, Box 351580, Seattle, WA 98195, USA}

\author[0000-0002-3038-3896]{Zhuo Chen}
\affiliation{Department of Astronomy, University of Washington, Box 351580, Seattle, WA 98195, USA}

\author[0000-0003-3252-352X]{Margaret Lazzarini}
\affiliation{Department of Physics and Astronomy, California State University, Los Angeles, CA 90032, USA}

\author[0000-0002-1264-2006]{Julianne J. Dalcanton}
\affiliation{Center for Computational Astrophysics, Flatiron Institute, 162 Fifth Avenue, New York, NY 10010, USA}
\affiliation{Department of Astronomy, University of Washington, Box 351580, Seattle, WA 98195, USA}

\author[0000-0002-5564-9873]{Eric F. Bell}
\affiliation{Department of Astronomy, University of Michigan, Ann Arbor, MI 48109-1107, USA}

\author[0000-0002-7743-9906]{Kameron Goold}
\affiliation{Department of Physics and Astronomy, University of Utah, Salt Lake City, UT 84112, USA}

\author[0000-0001-8416-4093]{Andrew Dolphin}
\affiliation{Raytheon, Tucson, AZ 85756, USA}
\affiliation{Steward Observatory, University of Arizona, Tucson, AZ 85726, USA}

\author[0000-0001-7531-9815]{Meredith J. Durbin}
\affiliation{Department of Astronomy, University of California Berkeley, Berkeley, CA 94720, USA}
\affiliation{Department of Astronomy, University of Washington, Box 351580, Seattle, WA 98195, USA}

\author[0009-0006-8134-0497]{Stefany L. Fabian Dubón}
\affiliation{Graduate Center, City University of New York, 365 5th Avenue, New York, NY 10016, USA}

\author[0000-0003-0394-8377]{Karoline M. Gilbert}
\affiliation{Space Telescope Science Institute, 3700 San Martin Drive, Baltimore, MD, 21218, USA}
\affiliation{The William H. Miller III Department of Physics \& Astronomy, Bloomberg Center for Physics and Astronomy, Johns Hopkins University, 3400 N. Charles Street, Baltimore, MD 21218, USA}

\author[0000-0001-8867-4234]{Puragra Guhathakurta}
\affiliation{Department of Astronomy and Astrophysics, University of California Santa Cruz, University of California Observatories, 1156 High Street, Santa Cruz, CA 95064, USA}

\author[0000-0002-2165-5044]{Francois Hammer}
\affiliation{LIRA, Observatoire de Paris, PSL University, CNRS, Place jules Janssen, 92195 Meudon, France}

\author[0000-0001-6421-0953]{L. Clifton Johnson}
\affiliation{Center for Interdisciplinary Exploration and Research in Astrophysics (CIERA) and Department of Physics and Astronomy, Northwestern University, 1800 Sherman Ave., Evanston, IL 60201, USA}

\author[0000-0001-9605-780X]{Eric W. Koch}
\affiliation{National Radio Astronomy Observatory, 800 Bradbury SE, Suite 235, Albuquerque, NM 87106 USA}

\author[0000-0001-5538-2614]{Kristen.~B.~W. McQuinn}
\affiliation{Space Telescope Science Institute, 3700 San Martin Drive, Baltimore, MD, 21218, USA}
\affiliation{Rutgers University, Department of Physics and Astronomy, 136 Frelinghuysen Road, Piscataway, NJ 08854, USA} 

\author[0000-0002-9820-1219]{Ekta~Patel}
\affiliation{Department of Astrophysics and Planetary Sciences, Villanova University,  800 E. Lancaster Ave, Villanova, PA 19085, USA}

\author[0000-0002-8406-0136]{Vaishnav V. Rao}
\affiliation{Department of Astronomy, University of Michigan, Ann Arbor, MI 48109-1107, USA}

\author[0000-0001-6326-7069]{Julia Roman-Duval}
\affiliation{Space Telescope Science Institute, 3700 San Martin Drive, Baltimore, MD, 21218, USA}

\author[0000-0003-2599-7524]{Adam Smercina}\thanks{NHFP Hubble Fellow}
\affiliation{Space Telescope Science Institute, 3700 San Martin Drive, Baltimore, MD, 21218, USA}


\author[0000-0002-6440-1087]{Debby Tran}
\affiliation{Department of Astronomy, University of Washington, Box 351580, Seattle, WA 98195, USA}

\author[0000-0002-6442-6030]{Daniel R. Weisz}
\affiliation{Department of Astronomy, University of California Berkeley, Berkeley, CA 94720, USA}


\correspondingauthor{Tobin M. Wainer}
\email{tobinw@uw.edu}

\begin{abstract}
We use \textit{Hubble Space Telescope} optical imaging from the Panchromatic Hubble Andromeda Southern Treasury (PHAST) to measure the spatially resolved recent star formation history (SFH) across the southern disk of M31. We fit color–magnitude diagrams (CMDs) of over 6500 individual 0.01 kpc$^2$ regions to measure SFHs over the last $\sim 500$ Myr. The resulting maps show coherent structure which trace the ringed morphology of the disk. We find a clear global decline in the recent star formation rate (SFR), with a pronounced drop in the last $\sim 40$ Myr that is most evident in the region closest to M32. 
Combining PHAST and PHAT measurements, we now cover two-thirds of M31's star forming disk with homogeneous SFHs derived with the same technique, yielding the highest resolution spatially resolved SFHs of M31 measured with resolved stars. Inside the joint footprint, we measure a mean SFR of $0.445 \pm 0.006\ \rm{M_\odot\ yr^{-1}}$ over the last 100 Myr, and $0.285 \pm 0.014\ \rm{M_\odot\ yr^{-1}}$ over the last 20 Myr, implying a total disk SFR of $\sim 0.67\ \rm{M_\odot\ yr^{-1}}$ and $\sim 0.43\ \rm{M_\odot\ yr^{-1}}$ respectively when extrapolated to the full star–forming disk. The observed decline is interpreted to be the late stages of a multi-Gyr wind-down from a previously more active state. Because recent star formation in M31 is concentrated primarily in the rings, the observed global decline is driven mainly by decreasing activity within those features. We also compare the CMD-based SFR surface densities to those inferred from FUV+24 \micron\ prescriptions and find that the FUV-based calibration underestimates the CMD–based 100 Myr average by a factor of $\sim 2.1$. However, the PHAST SFHs produce a synthetic \emph{GALEX} FUV image that agrees well with observations, indicating that the CMD–derived SFHs provide an accurate description of recent star formation. The mismatch with the FUV+24 $\micron$ estimates underscores that tracers implicitly averaged over $\sim100$ Myr are not reliable when the recent SFR is evolving.
\end{abstract}


\section{Introduction}

One key tracer of a galaxy's formation and evolution is its star formation history (SFH). This simple metric of how much stellar mass has formed each year constrains the galaxy's mass assembly and chemical evolution, offering clues about the astrophysical phenomena that have influenced its evolution \citep[e.g.,][]{holtzman_observations_1999, skillman_star_2005, tolstoy_star-formation_2009, conroy_modeling_2013, weisz_star_2014, lewis_panchromatic_2015, lazzarini_panchromatic_2022}. Focusing on the recent (<1 Gyr) SFH, we can directly measure the connection between local star formation events and the surrounding dust and gas, allowing us to study the physics of star formation and feedback \citep[e.g.,][]{cannon_m81_2011, christensen_environment_2024}. However, this detailed level of analysis can only be done for nearby galaxies with resolved stellar photometry, where we can spatially resolve the SFH for a specific region using samples of the stars that were formed \citep[e.g.,][]{mcquinn_nature_2012,lewis_panchromatic_2015, lewis_panchromatic_2017}. This distance limitation is particularly acute for massive galaxies, due to their high intrinsic stellar density.

A precise method that can be used to measure SFHs with the highest time resolution is color magnitude diagram (CMD) fitting. In this method, the number of stars of various ages and masses is counted directly, and the distribution of stars on the CMD is used to infer time-resolved star formation rates (SFRs) that would produce the observed population \citep[e.g.,][]{dolphin_numerical_2002, weisz_recent_2008, williams_history_2011}. A spatially-resolved SFH can be constructed by using CMD-fitting to measure the SFH independently in multiple spatial regions across a galaxy, which can then be tiled together to form a cohesive map of recent star formation \citep{lewis_panchromatic_2015, lazzarini_panchromatic_2022, tran_spatially_2023}.

Like all methods used to measure a galaxy's SFH, CMD analysis relies on assumptions about stellar evolution, binary populations, the initial mass function (IMF), and the distribution and amount of dust \citep{dolphin_estimation_2013}. CMD derived SFHs also have better time resolution at younger ages opposed to older ages. Small changes in stellar population age produce large changes in the color-magnitude distribution of stars, particularly in the well-populated, well-understood main sequence. In contrast, older populations are constrained more heavily by evolved phases such as the red clump and horizontal branch, where the age sensitivity is weaker and the modeling is less certain \citep[e.g.,][]{dolphin_numerical_2002,gallart_adequacy_2005,tolstoy_star-formation_2009,dib_assessing_2025}. 

CMD-based spatially resolved SFHs have been wildly successful in mapping nearby galaxies, which offer the most dynamic range of age sensitive CMD features. Examples include the SMC \citep{harris_star_2004, cohen_scyllaIII_2024}, LMC \citep{harris_star_2009, cohen_scyllaII_2024}, M33 \citep{lazzarini_panchromatic_2022}, and M31 \citep{williams_recent_2003,lewis_panchromatic_2015}; as well as more distant systems such as M81 \citep{choi_testing_2015}, NGC 300 \citep{gogarten_advanced_2010}, NGC 2403 \citep{williams_acs_2013}, and NGC 6946 \citep{tran_spatially_2023}.

M31 is an especially powerful laboratory for this kind of work. At a distance assessed to be 785 kpc \citep{mcconnachie_distances_2005}, M31 is the nearest massive L$_{*}$ spiral that can be resolved into individual stars. As a result, its recent SFH has been the focus of numerous resolved-star studies in selected fields \citep{williams_clues_2002,williams_recent_2003, brown_detailed_2006, brown_extended_2007, brown_extended_2008, bernard_star_2012, davidge_recent_2012}. These studies established that M31 has a complex and spatially varying SFH, most notably that the galaxy appears to have a lower SFR at more recent look-back times, compared to times larger than 1 Gyr. However, these studies also reflected the usual trade-off between photometric depth, which improves age resolution and helps reduce age–metallicity degeneracies \citep[e.g.,][]{gallart_adequacy_2005}, and spatial coverage. Thus, early studies had limited spatial coverage and heterogeneous sampling of the disk, making it difficult to connect temporal variations in the SFR to galaxy-wide structure.


However, M31's deep photometric coverage was greatly improved when a significant portion of the galaxy was observed with \textit{HST} as a part of the Panchromatic Hubble Andromeda Treasury (PHAT) \citep{dalcanton_panchromatic_2012}. PHAT spans the near-UV through the near-IR, providing resolved stellar photometry for over 100 million stars. The survey's combination of depth and contiguous spatial coverage made it possible to connect M31's local stellar populations to larger-scale disk structure in a way that had not been feasible in earlier targeted-field studies.

One notable study which used these data is that of \citet[][hereafter L15]{lewis_panchromatic_2015}, which provided the first spatially resolved SFH for a significant portion of M31. \citetalias{lewis_panchromatic_2015} divided the 0.5 square degree area in the northern third of the disk covered by PHAT into roughly 9000 regions measuring 100 pc by 100 pc. They optimized their fits for recent star formation, accurate within the last $\sim 500$ Myr, focused on reconstructing the blue half of the CMD, which is populated almost entirely by young main sequence stars. 

The results from \citetalias{lewis_panchromatic_2015} have been incredibly valuable in precisely characterizing M31. For galactic structure and evolution, the \citetalias{lewis_panchromatic_2015} SFH maps have helped establish how long-lived the 10 kpc ring is, and that the PHAT region in M31 had a modest burst of star formation $\sim100$ Myr ago, in addition to the large burst seen $\sim2$ Gyr ago \citep{williams_panchromatic_2014} most likely in the aftermath of a recent merger \citep{dsouza_andromeda_2018, hammer_2-3_2018}. 

The maps have also been used to better calibrate various SFR indicators, via direct comparison with the CMD-based SFRs \citep{lewis_panchromatic_2017}. For stellar astrophysics, \citet{diaz-rodriguez_progenitor_2018} used the \citetalias{lewis_panchromatic_2015} SFH maps to measure the progenitor mass distribution for core collapse supernovae. Additionally, the maps have been used to measure the age distribution of high-mass X-ray binaries \citep{lazzarini_young_2018, williams_comparing_2018, lazzarini_multiwavelength_2021}. For gas and dust studies, the \citetalias{lewis_panchromatic_2015} maps are an important indicator of local densities of young stars \citep[e.g.,][]{lindberg_dust_2024, eknath_adventures_2024}. They also provide a key observational constraint for theoretical simulations of galactic dynamics \citep[e.g.,][]{feng_bar-driven_2024, hammer_dark_2025}. 

In contrast to the well-studied region of M31 covered by the PHAT survey, the southern disk has not yet been as well characterized, despite being notably different and more complex than the northern half. The largest, and arguably the most interesting, feature distinguishing the southern half of the disk is the presence of M32, just 5 kpc from the center of the M31 disk in projection, although with unknown distance to the disk itself \citep[e.g.,][]{mcconnachie_observed_2012}. Due in part to this distance uncertainty, and equally large kinematic uncertainties, the exact history of interaction between M32 and M31 is unknown. Some models suggest that M32 is the stripped remnant of a recent merger \citep{bekki_new_2001,  dierickx14,dsouza_andromeda_2018, escala_kinematical_2025}, and the prevailing theory is that M32 is currently impacting the M31 disk \citep[e.g.,][]{gordon_spitzer_2006,patel_m31m33_2025}. 

Whether due specifically to M32 or otherwise, the southern half of the M31 disk appears to be more perturbed by its merger history than the northern half \citep{gordon_spitzer_2006, dsouza_andromeda_2018, hammer_2-3_2018}. Additional unique features in the southern half of the M31 disk include the giant southern stream \citep{ibata_giant_2001, ibata_haunted_2007}, and NGC~206, M31's largest star forming region \citep{brinks_ngc_1981, odewahn_photometric_1987}. The galaxy's largest stellar clusters are in the south \citep{perina_hstwfpc2_2009}, which could be evidence of a higher star formation rate surface density ($\Sigma_{SFR}$) \citep{johnson_panchromatic_2017, wainer_panchromatic_2022}.

Each of these features may either influence the local SFH or reflect a distinct evolutionary history relative to the northern disk. Thus, the southern disk is a natural target for a detailed modeling of the recent SFH, which could yield valuable clues to the relationship between M31 and M32, and link the prominent structures seen today to M31's global evolutionary history. Spatially resolving the SFH makes that connection especially meaningful by tracing these southern structures in both space and time.

The opportunity to study the southern disk with deep, high precision photometry is now available with the Panchromatic Hubble Andromeda Southern Treasury (PHAST) survey \citep{chen_phast_2025}, obtained using the \textit{Hubble Space Telescope} (HST). With over 90 million resolved stars covering 0.45 deg$^2$ of the star forming disk, PHAST presents reliable optical photometry down to $\sim27$ magnitude corresponding to an absolute magnitude of $\sim2.5$. This depth is more than adequate for precisely measuring the star formation activity over the past 500 Myr in M31 with excellent time resolution (see Section 3.5 for more details). With this comprehensive data set, we seek to follow the methodology of \citetalias{lewis_panchromatic_2015} and measure the spatially resolved SFH within the PHAST footprint, using CMD fitting to describe the SFH in $100\times100$ pc regions. 


By combining the PHAT and PHAST surveys, we gain access to resolved-stellar photometry for roughly 0.2 billion stars. Together, these homogeneous \emph{HST} data sets provide contiguous coverage of roughly two-thirds of the star-forming disk, enabling the highest-resolution galaxy-scale SFH map of M31 yet measured from resolved stars, and the only such detailed map of any L$_{*}$ galaxy. Combining the SFH maps created with both the PHAT \citepalias{lewis_panchromatic_2015} and PHAST surveys enables the first truly galaxy-scale analysis of the recent SFH in M31. With this combined coverage, we can directly test asymmetries between the northern and southern disk, connect local star-forming structures to global trends, and trace how star formation has evolved across M31 over the last $\sim 500$ Myr.

This paper is structured as follows. We describe the PHAST data in Section~\ref{sec:data} and describe the CMD based SFH fitting in Section~\ref{sec:methods}. We present the SFH for PHAST in Section~\ref{sec: results}, and combine the PHAT and PHAST results in Section~\ref{sec:PHAT_PHAST_combined}. We then compare the SFRs derived from our CMD based analysis to other indicators in Section~\ref{sec:other_tracers}. Finally, we compare the northern and southern disk, and discuss the M31 SFH in the context of galaxy evolution in Section~\ref{sec: discusion}, including consideration of specific sub-regions such as the vicinity of M32, and the inner arm, in Section~\ref{sec:sub_regions}. 

\section{PHAST Data}\label{sec:data} 

Our analysis is based on imaging from the PHAST survey \citep{chen_phast_2025}, a large HST program targeting the southern disk of M31 (GO-16778 and 16796–16801; PI: Williams). A full description of the survey, along with figures of the survey strategy can be found in \citet{chen_phast_2025}, a brief summary follows.

PHAST builds upon the strategy of the PHAT \citep{dalcanton_panchromatic_2012} and PHATTER \citep{williams_panchromatic_2021} surveys, providing contiguous four-band coverage in the F275W, F336W, F475W, and F814W filters using single-orbit observations from ACS/WFC and WFC3/UVIS. Our work here only uses the optical F475W and F814W wavelengths due to their depth and our focus on young stars, as discussed more in Section~\ref{sec:methods}. The survey extends $\sim$13~kpc along the southern major axis of M31, covering a total area of $\sim$1640 arcmin$^2$.

Observations were conducted at two orientations separated by 180$^\circ$, which placed the parallel ACS/WFC exposures over the location of primary WFC3/UVIS exposures. Each pointing received one orbit of exposure time, with two dithered exposures in each filter to mitigate detector artifacts and improve sampling. The footprint of PHAST, combined with PHAT, covers approximately two-thirds of the far-ultraviolet (FUV) flux from M31’s disk, based on comparison with \emph{GALEX} imaging \citep{martin_galaxy_2005}. 

\subsection{Photometry and Catalogs}

Stellar photometry was measured using the \texttt{DOLPHOT} package \citep{dolphin_wfpc2_2000, dolphin_dolphot_2016}, following procedures established in PHAT and PHATTER. For a full description of the intricacies of deriving the PHAST photometry, refer to \citet{chen_phast_2025}. Here we provide a brief synopsis. 

Photometry was carried out using PSF fitting across stacked subregions containing all overlapping exposures. The photometry pipeline incorporates cosmic-ray masking, pixel area maps, PSF and aperture corrections, and produces calibrated magnitudes for all stars detected in any band. Each subregion includes overlapping exposures from adjacent bricks and fields, ensuring consistent depth and reducing edge artifacts. The final PHAST catalog contains $\sim$91 million sources with full 4-band photometry, signal to noise estimates, crowding and sharpness diagnostics, and quality flags. We use the \texttt{gst} subset of the catalog, which includes only stars with reliable photometry in each band (S/N $>$ 4, sharpness$^2 < 0.2$, and crowding $< 2.25$). This catalog forms the basis for CMD fitting in this work.

\subsection{Region Size and Stellar Density Determination} \label{sec:region_size}

Each region in our spatial grid spans $\sim$26\arcsec per side, binned by RA and Dec. At the distance of M31 at 785 kpc, this corresponds to square pixels that equate to 100 pc per side. However, due to the inclination of the galaxy ($i=77\degree$; \citealt{walterbos_optical_1988}), projection makes each region larger by a factor of $1/$cos($i$). Throughout the rest of this work, when we consider the area of each region, we refer to the de-projected area, or 0.0445 kpc$^2$. This region size balances the need for resolved structures (e.g., OB associations, H~\textsc{ii} regions) with sufficient CMD sampling to constrain the SFH \citep[e.g.,][]{simones_panchromatic_2014}. 

\begin{figure*}[ht]
    \centering    
    \includegraphics[width=0.95\textwidth]{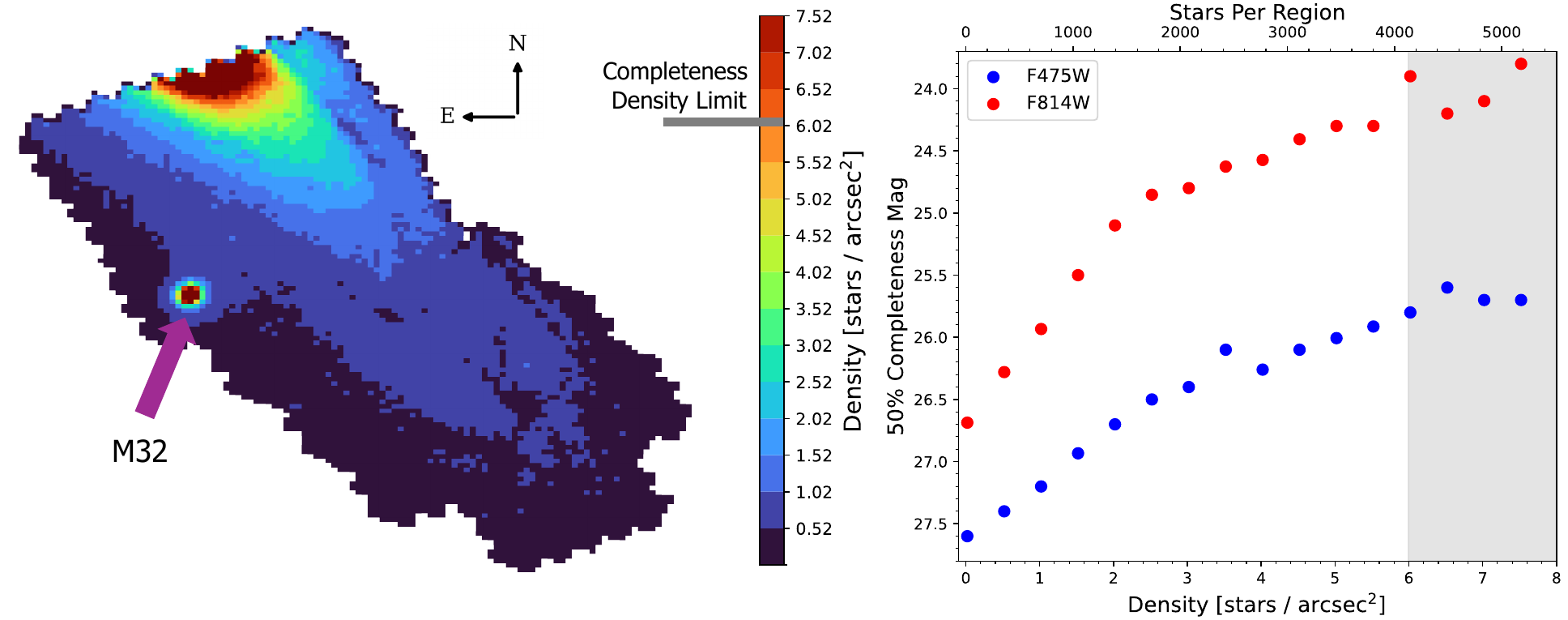}
    \caption{\textit{Left:} Stellar density in the PHAST survey, as defined by stars per square arcsecond with 21.5 < F814W < 23. \textit{Right:} 50\% completeness limits as determined by ASTs as a function of local stellar density, estimated from bootstrapped sampling of across the PHAST footprint. Completeness declines at higher stellar densities due to crowding. The gray shaded region in the right panel, and illustrated on the color bar of the left panel, indicates} regions not included in the current study (Section~\ref{sec:age_range_and_reliability}).
    \label{fig:completeness_vs_density}
\end{figure*}

We characterize the stellar density following the \citet{chen_phast_2025} prescription. Specifically, stellar density is determined by the number of \texttt{gst} stars per square arcsecond with $21.5 < \rm{F814W} < 23$. This density across the footprint is shown in the left panel of Figure~\ref{fig:completeness_vs_density}, where density is binned by steps of 0.5 stars per arcsec$^{2}$. As expected, there is a clear gradient of decreasing density moving from the dense center of the galaxy down and to the right along the disk. Notably however, there is a very prominent dense region near the survey edge, which is M32.

We note that in our analysis we only includes regions where that are fully covered by the PHAST imaging. Because our grid was done in RA and Dec, any regions that lie on the perimeter of the survey only have partial coverage, and therefore incorrect density estimates, and are excluded. 

\subsection{Artificial Star Tests}\label{sec:asts}

A standard way to characterize photometric uncertainties, completeness, and bias due to crowding, is to use artificial star tests (ASTs) inserted throughout the survey area \citep{dolphin_numerical_2002,dalcanton_panchromatic_2012,williams_panchromatic_2021,chen_phast_2025}. We follow this common approach, inserting fake stars to survey fields individually, covering a range of magnitudes and colors, and processing the image through the same photometric pipeline as the original imaging data to assess recovery. To quantify this dependence, we compute completeness curves using ASTs drawn from bins in stellar density, and identify the magnitude at which 50\% of inserted stars are recovered within 1~mag and 2 pixels (See Figure~\ref{fig:completeness_vs_density}).

In the PHAST survey \citep{chen_phast_2025}, five regions along the southern major axis of the disk were initially selected for ASTs. These regions were chosen to span the full dynamic range of Red Giant Branch (RGB) stellar densities observed throughout the survey area. To refine our completeness model for the current work, we expanded this sample by conducting identical AST procedures on seven additional sub-regions chosen to specifically supplement areas of density space. 

Our AST library therefore spans twelve representative fields, and creates a high-fidelity lookup foundation from which the subsequent density-dependent sampling can be confidently performed for any observed region on the disk. Together, this constitutes a library of 670,000 ASTs in total, each with an associated local density, providing the necessary empirical framework for the SFH modeling described in this work.

We adopt the 50\% completeness limit of our ASTs as the faintest magnitude used in our CMD fitting. These limits vary across the survey as a function of local stellar density, with denser regions towards the center of the galaxy having brighter 50\% completeness limits, as seen in the right panel of Figure~\ref{fig:completeness_vs_density}. We choose bin widths of 0.5 stars per square arcsec, sampled from densities of 0 to 8 stars per square arcsec, and randomly select 50,000 ASTs for each density bin. To better reflect the CMD structure of each region, we supplement the straight density-based AST selection with an additional subset drawn from the magnitudes ranges specifically populated in observed CMD morphology (i.e., we draw 15,000 more ASTs for the color and magnitude ranges where we explicitly observe stars in each region). For each density bin, we repeat the completeness calculation 15 times using bootstrap resampling of 65,000 ASTs per trial, and adopt the mean 50\% completeness magnitude in F475W and F814W. The right panel of Figure~\ref{fig:completeness_vs_density} shows the median 50\% completeness magnitude. We can see a clear trend in 50\% completeness magnitude as a function of stellar density. These 50\% completeness magnitudes correspond to specific ages for the main sequence turnoff, which is visualized in Figure~\ref{fig:turnoff}, and discussed further in Section~\ref{sec:age_range_and_reliability}.

To test for any bias introduced by the selection of ASTs we performed a series of sensitivity tests by assessing how varying choices for our standard procedure affect the measured 50\% completeness limits, and subsequent SFH. We describe these tests below, and for each test, we found that the SFH changed by less than the $1\sigma$ uncertainties. Specifically, we: 1) increase the width of density bins we draw the AST from to be within 0.5, and 1 stars per square arcsec, 2) only select ASTs entirely from the density bin higher and lower than the current region, and 3) change the number of ASTs to range from 50,000 to 100,000. In each case, the resulting SFH varied by much less than the 1$\sigma$ uncertainty. These tests suggest the 65,000 ASTs well sample the population within the region, and the SFH is robust to minor fluctuations in stellar density. Additionally, while the $50\%$ completeness curve is not perfectly smooth between density bins, these small changes do not impact the SFH results. 

\section{Methods}
\label{sec:methods}

We derive spatially resolved recent SFHs across the PHAST footprint using the \texttt{MATCH} CMD-fitting package \citep{dolphin_numerical_2002}, following the established approach introduced by \citetalias{lewis_panchromatic_2015} and later applied across a range of galaxies \citep{lazzarini_panchromatic_2022, tran_spatially_2023}. A detailed description of the methodology can be found in \citetalias{lewis_panchromatic_2015}, while we provide a brief overview here.

\subsection{Derivation of the SFHs}\label{sec:derivation_of_sfh}

We reconstruct the time-dependent star formation rate for a given region by statistically comparing observed optical CMDs to synthetic CMDs generated from stellar evolution models, convolved with photometric uncertainties and completeness from artificial star tests (ASTs). The likelihood of each model is computed using a Poisson maximum likelihood statistic, and the best-fit SFH corresponds to the model CMD with the highest likelihood.

Our analysis uses HST photometry in the F475W and F814W filters (Section~\ref{sec:data}), which provides the deepest coverage and the greatest leverage on recent star formation. We focus on the young stellar populations, following the practices of \citetalias{lewis_panchromatic_2015} and \citet{lazzarini_panchromatic_2022}, by excluding regions of the CMD dominated by older stars by removing sources with \mbox{F475W$-$F814W~$>$~1.25} and \mbox{F475W~$>$~21}, a color-magnitude cut that minimizes contamination from the RGB and red clump while preserving reddened massive stars and helium-burning phases. This exclusion region is shown as a gray region outlined by a dashed box when we plot CMDs (see Figures~\ref{fig:turnoff} and Figure~\ref{fig:showing_diff_regions}). These limits include young red and blue He burners in our fits which are useful for age dating at young ages \citep[e.g.,][]{dohm-palmer_dwarf_1997,mcquinn_nature_2012}.

We fit the CMD using the following choices for \texttt{MATCH}. The stellar population models are drawn from the Padova 2006 isochrones \citep{marigo_evolution_2008} with CO ratio support \citep{girardi_acs_2010}, and the \citet{girardi_red_2016} translations. We assume a Kroupa IMF \citep{kroupa_variation_2001}, a binary fraction of 0.35 with uniformly distributed secondary masses, and a fixed distance modulus of 24.47. We assume metallicity increases with time, with [M/H] ranging from $-2.3$ to $+0.1$. The SFH is solved in 34 logarithmic time bins from log(age/yr) = 6.6 to 10.15, with 0.1 dex spacing except for the final bin which spans log(age/yr) = 9.9-10.15.

\subsection{Correcting for IMF Integration Limits}
\label{sec:imf_limits}

The SFHs used throughout this work were derived with \texttt{MATCH}, assuming a \citet{kroupa_variation_2001} IMF. As noted by \citet{telford_mass--light_2020}, \texttt{MATCH} normalizes the IMF by integrating over the full mass range from $0$ to $\infty$, rather than over the more commonly adopted finite stellar-mass limits of $0.1$–$100~M_\odot$. This choice does not affect the relative distribution of stars across the CMD, but it does affect the conversion between the fitted number of stars and the total stellar mass formed. As a result, the SFRs reported directly by \texttt{MATCH} are systematically offset from those corresponding to the same IMF evaluated over finite stellar-mass limits.

Using the same IMF slopes adopted in the \texttt{MATCH} fits, we compute the ratio between the IMF normalization over $0.1$–$100M_\odot$ and that obtained over the effectively unbounded mass range assumed internally by \texttt{MATCH}. We then apply a uniform correction to all reported SFRs so that they correspond to a Kroupa IMF integrated from $0.1$ to $100M_\odot$.

This correction yields a multiplicative correction factor of $f_{\rm IMF} = 0.799$, corresponding to a -20.096\% decrease in all SFR values. We apply this correction uniformly to every SFH time bin and to all derived SFR quantities reported in this paper. This correction only changes the absolute normalization of the SFHs, and does not affect any of the relative temporal or spatial trends discussed below.

\subsection{Extinction}
The stars within any given PHAST region are subject to a range of line–of–sight extinctions, which leads to differential attenuation imprinted directly on the morphology of the observed CMD. To recover reliable recent SFHs, we must therefore model the impact of both foreground and internal dust. 

When modeling the effects of dust on the CMD, foreground dust is treated as a screen with total extinction $A_V$ that affects all stars equally, shifting all features to redder colors and fainter magnitudes. Internal dust is fundamentally more complicated, as it depends on the details of the star-gas geometry, which can be different depending on the ages and locations of a particular stellar sub-population. For the young stars that we focus on here, we adopt a ``uniform slab'' model where young stars are assumed to be evenly mixed within a uniform layer of dust of total extinction $dA_V$, such that some stars near the front of the slab have little additional extinction, and those near the back experience the full $dA_V$. This two component $A_V+dA_V$ model is flexible enough to reproduce the observed broadening and dimming of stars across the CMD.
 
Specifically, our adopted model of dust attenuation has two free parameters per region: a uniform extinction $A_V$ to model for foreground dust, and a differential extinction component $dA_V$, which introduces a spread in extinction values across the stellar population, equivalent to assuming the stars are uniformly distributed within a layer of dust. The foreground extinction, $A_V$, shifts the CMD redward and fainter, while the differential component, $dA_V$, broadens and differentially dims the main sequence, but keeps the blue edge fixed. This approach follows the implementation in \citetalias{lewis_panchromatic_2015}, where the total extinction is applied as a tophat distribution between $A_V$ and $A_V + dA_V$. 

This extinction model is appropriate for young stellar populations (ages $\lesssim 1$~Gyr), which are confined to the disk and tend to experience relatively uniform extinction on the scales probed here \citep{dolphin_deep_2003, weisz_star_2014}. In \texttt{MATCH}, the application of $A_V$ and $dA_V$ in a uniform distribution across all stars provides an effective first-order treatment of reddening for young stars, though notable departures are seen for older stellar populations that are not confined to the scale height of the dusty ISM \citep[e.g.,][]{dalcanton_panchromatic_2015, choi_smashing_2018}.

We determine the best-fit extinction values by evaluating the SFH over a grid of $A_V$ and $dA_V$ combinations. We initially sample $A_V$ in the range [0.0, 1.5] and $dA_V$ in the range [0.0, 2.5]\footnote{See Appendix B in \citetalias{lewis_panchromatic_2015} for a discussion on why these limits were chosen.} using a coarse grid with spacing of 0.2 in each parameter. We then refine the solution by constructing a finer grid with spacing of 0.05 around the best-fit ($A_V$, $dA_V$) pair, limited to the region within 2$\sigma$ of the minimum fit statistic. This two-step grid search allows us to efficiently locate the best-fitting extinction parameters while assessing uncertainty in the dust distribution.

\begin{figure*}[ht]
    \centering
    \includegraphics[width=0.95\textwidth]{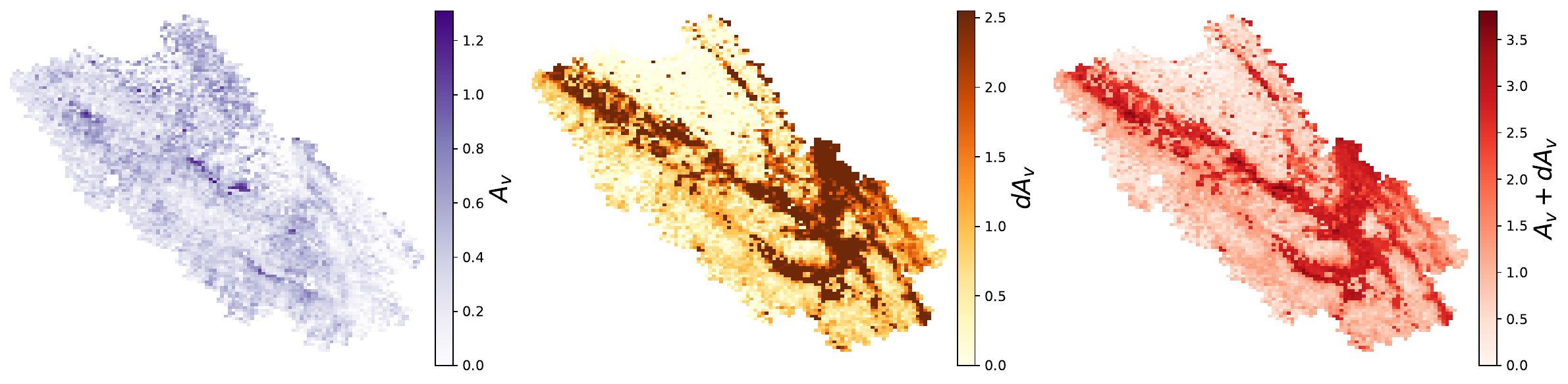}
    \caption{Spatial distribution of recovered extinction parameters across the PHAST footprint, as determined by the CMD-fitting procedure in each analysis region. Left panel shows the foreground extinction $A_V$, while the middle panel shows the differential extinction $dA_V$, and the right panel shows the total extinction $A_V + dA_V$. Regions are color-coded by extinction, with darker values indicating higher total extinction. Elevated extinction values trace prominent dust structures which correspond to arms and active star-forming regions (refer to color image in Figure~\ref{fig:showing_diff_regions}). The spatial coherence of this morphology provides confidence in the CMD-based recovery of dust attenuation across the disk.}
    \label{fig:av_dav}
\end{figure*}

We measure modest foreground extinction values across the PHAST regions, ranging from $A_V \approx 0.1$ mag up to 1.2 mag, consistent with the values from \citetalias{lewis_panchromatic_2015}. Each region's best fit $A_V$ is seen in the left panel of Figure~\ref{fig:av_dav}. While there is some slight structure in the $A_V$ map, these values are mostly uniform across the disk, representing the foreground extinction. The values recovered here are also broadly consistent with the \citet{schlegel_maps_1998} and \citet{schlafly_measuring_2011} E(B-V) maps in the regions near M31. However, some of the residual structure indicates regions where some of M31's complex gas distribution might create a ``foreground'' that is associated with M31, not the Milky Way. 

In contrast, our inferred differential extinction ($dA_V$) extends to higher values, demonstrating a wider dynamic range across the disk. Elevated $dA_V$ is frequently associated with dense star-forming structures, especially along spiral arms, as expected for an approximately uniform dust-to-gas ratio. When plotted spatially, the dust morphology, traced from the $A_V + dA_V$, maps out known galactic structure, reinforcing the reliability of our extinction mapping, which is done independently in each region, but produces coherent, physically sensible maps.

This spatial structure is shown in Figure~\ref{fig:av_dav}, which shows the best-fit $\rm{A_{V} + dA_V}$ values across the PHAST footprint. The recovered extinction morphology tracks expected features of M31’s disk, such as the outer spiral arms and active star-forming ring, with elevated $dA_V$ appearing in dense, or visible dusty regions. The overall structure, and dynamic range, of extinction across PHAST regions appear consistent with known patterns of star formation and dust. Moreover, visual inspection of CMDs and recovered SFHs in high-extinction regions reveals no signs of inadequate fits, suggesting the CMD-based extinction solutions are robust even in dusty regions.


\subsection{Uncertainties} \label{sec: uncertainty}

Uncertainties in our SFH measurements primarily arise from a combination of statistical, systematic, and extinction-related effects. We follow the uncertainty estimation procedures described in \citet{dolphin_estimation_2013}, adopting the same techniques used in previous applications of this method (\citetalias{lewis_panchromatic_2015}; \cite{lazzarini_panchromatic_2022, tran_spatially_2023}.

\subsubsection{Random Uncertainties}
Statistical uncertainties are estimated using the hybrid Markov Chain Monte Carlo (MCMC) routine built into \texttt{MATCH}. This sampler efficiently explores high-dimensional parameter space and generates 10,000 SFH realizations per region, with the density of samples proportional to the posterior probability. The 1$\sigma$ confidence interval is determined by identifying the region that contains 68\% of the realizations, even in bins where the best-fit SFR is zero.

Dust-related uncertainties are introduced by the degeneracy between reddening and stellar age, as well as our choice of spacing between ($A_V$, $dA_V$) grid points. To account for this uncertainty, we marginalize over all SFH solutions within 2$\sigma$ of the best-fit extinction parameters ($A_V$, $dA_V$), weighting each solution by the likelihood relative to the best fit. This weighting follows a Gaussian likelihood kernel (i.e., $e^{-0.5 n^2}$), where $n$ is the number of standard deviations above the minimum fit statistic.

We note that our assumed binary fraction (0.35) may be inconsistent with studies showing a dependence of multiplicity fraction on stellar mass \citep{offner_origin_2023}. In \texttt{MATCH}, the adopted binary fraction is not independent of the companion mass-ratio distribution, which is assumed to be uniform from 0 to 1. As a result, some differences in the true binary population could manifest as differences in the mass-ratio distribution rather than the overall binary fraction. In practice, previous tests show that reasonable variations in the adopted binary fraction have negligible impact on the recovered SFH compared to uncertainties from dust and photometric noise \citepalias{lewis_panchromatic_2015}.


\subsubsection{Systematic Uncertainties from Stellar Evolution Models}
\label{sec:isochrone_uncertainties}

In addition to the random and dust-related uncertainties discussed above, there is a further source of systematic uncertainty associated with our choice of stellar evolution models, and the fixed \citet{kroupa_variation_2001} IMF. We discuss both of these additional assumptions here. 

In \citet{lazzarini_panchromatic_2022}, a comparison between Padova and MIST isochrones for the full PHATTER survey showed that the two model sets generally agree within the measured uncertainties for most time bins older than $\sim 20$~Myr, while the largest discrepancies appear at the youngest ages. At ages $\lesssim 15$~Myr, they found that the inferred SFRs from different model sets can differ by up to a factor of two.

To assess the impact of these model choices in PHAST, we repeated the entire SFH fitting for a representative subset consisting of roughly a third of our regions using the MIST \citep{choi_mesa_2016} isochrones in place of the Padova \citep{marigo_evolution_2008, girardi_acs_2010} tracks. The comparison sample spans a wide range of stellar densities and recent SFRs, capturing the diversity of conditions present in the full survey. For each region in this subset we compared the recovered SFHs from the Padova and MIST fits in the same logarithmic time bins used throughout this work.

We find that for most of time bins in our analysis, the Padova and MIST SFHs agree to within the formal random uncertainties, and never exceed $2\sigma$ deviation. We note however, that the residuals are not symmetric. Across the sample, across all time bins the MIST models find a median $\sim30\%$ increase relative to the Padova models.

The other source of systematic uncertainty we consider is the choice of a fixed \citet{kroupa_variation_2001} IMF. There has been recent evidence to support that the high-mass IMF slope in nearby galaxies is steeper than this canonical value \citep{weisz_high-mass_2015, wainer_panchromatic_2024}. Specifically, using the PHAT data in M31, \citet{weisz_high-mass_2015} measured a high mass slope of $\Gamma = 1.45^{+0.03}_{-0.06}$. To determine the impact this difference has on our derived SFHs, we repeated the entire SFH fitting for the same representative subsample with the \citet{weisz_high-mass_2015} IMF slope. We find that the derived SFHs are equivalent within the uncertainties, with the higher IMF slope systematically recovering $\Sigma_{\rm SFR}$ values only $2\%$ higher than the baseline model, a much smaller deviation than that found between the stellar models. This result likely indicates that the systematic errors from stellar evolutionary models are less than the age-to-age differences in the recovered SFHs. 


We note that in both the IMF and isochrone model analysis, the spread in recovered $A_V + dA_V$ values is larger than the consistency in $\Sigma_{\rm SFR}$ might suggest. For the comparison between MIST and Padova, the standard deviation of the squared difference $\bigl[\,(A_V + dA_V)_{\rm Padova}-(A_V + dA_V)_{\rm MIST}\,\bigr]^2$ is 0.14 mag, and it is 0.17 mag for the IMF comparison. However, the best-fit likelihood values are extremely similar in both tests. This implies that differences between the model choices are largely accommodated by small shifts in the best-fit extinction parameters, with minimal change in the likelihood. Importantly, despite this flexibility in ($A_V, dA_V$), the recovered $\Sigma{\rm SFR}$ values remain statistically consistent between model choices, suggesting that the dominant effect of these differences is to adjust dust parameters, rather than to change the inferred recent star formation rate.

We do not fold this model dependence into the reported uncertainties on our SFHs, which represent the random uncertainties from the CMD fitting. Instead, we treat the model differences as an additional systematic uncertainty, particularly affecting the absolute normalization of the SFR. The detailed values of the SFR bins should therefore be interpreted with an additional slight systematic uncertainty of order $\sim10\%$, while the spatial patterns and longer-timescale trends remain robust.

\subsection{Reliable Age Range of SFH Measurements}\label{sec:age_range_and_reliability}

While the full \texttt{MATCH} time binning extends to several Gyr, the reliability of SFH recovery is fundamentally limited by the photometric depth in each region. Main-sequence turnoff stars dominate age sensitivity for young populations, but become increasingly faint and rare at older ages. Although useful age information can still be extracted from CMDs that do not reach the main-sequence turnoff, the resulting age constraints are generally less precise and less secure than those obtained from turnoff-resolved photometry.

Following \citetalias{lewis_panchromatic_2015}, we adopt a uniform 500~Myr lookback time as the threshold for reliable star formation constraints across the PHAST footprint. Imposing a consistent limit ensures spatial uniformity and avoids over-interpretation, though we note that low-density or dust-free regions can reliably probe somewhat older ages (Appendix~\ref{appendix}). 

\begin{figure}[ht]
    \centering
    \includegraphics[width=0.47\textwidth]{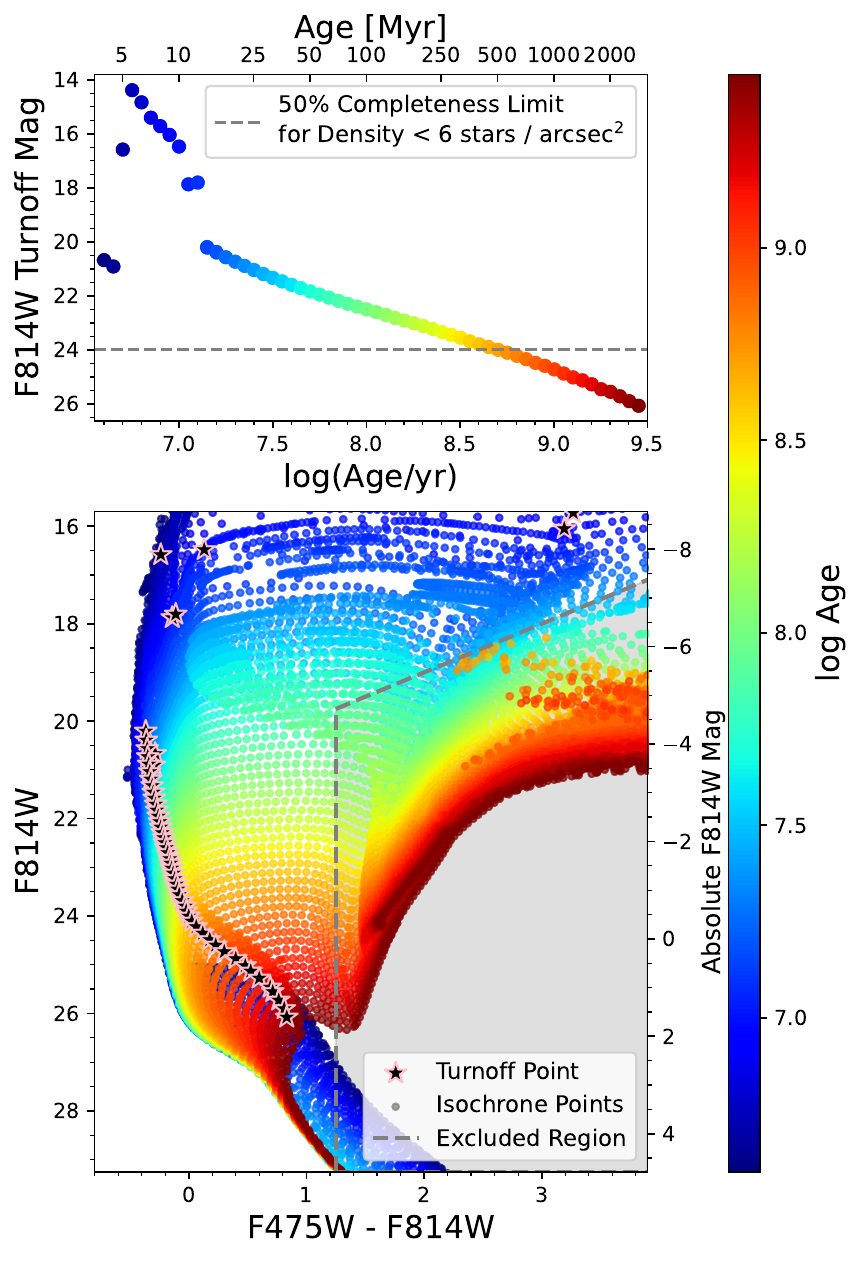}
    \caption{
    Padova stellar isochrones spanning log(Age/yr) = 6.6–10.2 plotted in CMD space (F475W$-$F814W vs. F814W), color-coded by age. Black stars mark the main-sequence turnoff point for each isochrone, defined as the transition from core hydrogen burning to the subgiant phase. The gray shaded region shows the excluded region described in Section~\ref{sec:derivation_of_sfh}. The secondary axis to the right shows the absolute F814 magnitude.} The top panel shows the F814W magnitude of the turnoff as a function of age, with a horizontal dashed line at F814W = 24~mag indicating the shallowest 50\% completeness limit across the PHAST survey. This provides a basis for our adopted 500~Myr threshold for reliable SFH measurements.
    \label{fig:turnoff}
\end{figure}

To empirically justify this threshold, we use the Padova stellar evolution models to trace the main-sequence turnoff as a function of age. For each isochrone in our grid, we consider the transition point from the hydrogen-burning main sequence to the subgiant branch from the model, and extract its F814W magnitude. Figure~\ref{fig:turnoff} shows the full set of isochrones in CMD space, color-coded by age, with black stars marking the turnoff points. 

The top panel summarizes the turnoff magnitude as a function of age, with a horizontal line indicating F814W = 24 mag which corresponds to the turnoff for the log(Age)=8.7 ($\sim500$ Myr) isochrone. From Figure~\ref{fig:completeness_vs_density}, we can see that regions with densities smaller than $\sim6$ stars per arcsec$^2$ have F814W 50\% completeness limits fainter than 24th mag. 

We therefore adopt 500 Myr as a conservative and physically motivated boundary for the time range of reliable SFH recovery, and choose to only consider regions where the stellar density is below 6 stars per arcsec$^2$. From Figure~\ref{fig:completeness_vs_density}, we can see that this density threshold only cuts off the very center of the galaxy (r $<\sim1$ kpc), and M32.

\section{PHAST Star Formation Histories}\label{sec: results}

In this section, we present the integrated and spatially resolved SFH for the PHAST survey. Full SFH results for each spatial region are presented in Table~\ref{tab:phast_sfh} \footnote{We note there are 15 spatial regions towards the very center of the galaxy that exceeded the available compute resources due to density and increased extinction. For these regions we note a -99 in the $A_V$, $dA_V$ and fitvalue columns of Table~\ref{tab:phast_sfh}. We mask these regions in our analysis, and fill in 0 SFR in the spatial maps.}. 

\begin{figure*}
    \centering
    \includegraphics[width=0.975\textwidth]{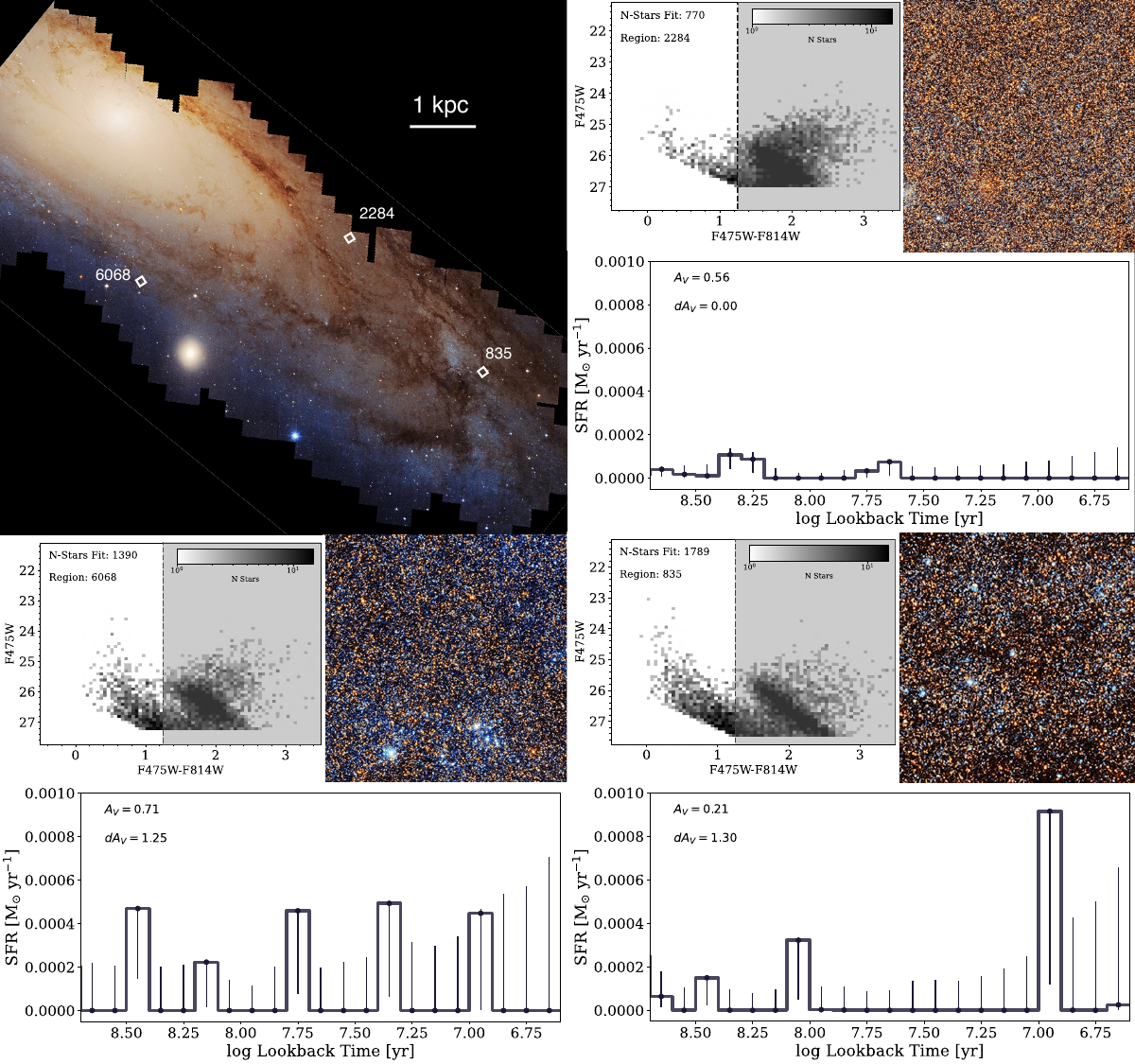}
    \caption{Overview of three representative PHAST regions spanning a range of stellar densities and environments.  
    \textit{Upper Left}: A color mosaic of the southern disk of M31, and the three selected regions (6068, 835, and 2284) highlighted (not perfectly to scale). For each region, we show: (upper left) the observed CMD used in SFH fitting, with excluded areas shaded in gray; (upper right) a color cutout of the region from the mosaic; and (bottom) the derived recent SFH. 
    These examples highlight the diversity of star-forming environments in the PHAST footprint and demonstrate the robustness of our CMD-fitting methodology across varying crowding and extinction conditions.}
    \label{fig:showing_diff_regions}
\end{figure*}

Figure~\ref{fig:showing_diff_regions} showcases three representative regions that illustrate the range of stellar densities and environments captured in the survey. The top left shows a color mosaic\footnote{Image credit: NASA, ESA, Benjamin F. Williams (University of Washington), Zhuo Chen (University of Washington), L. Clifton Johnson (Northwestern); Image Processing: Joseph DePasquale (STScI).} of the southern M31 disk with PHAST coverage, highlighting Regions 6068, 835, and 2284. These regions span a range of stellar densities from sparse outskirts to more crowded inner disk structures. Region 6068 lies near M32 and the 10 kpc ring, Region 835 sits along a well-defined dust lane, and Region 2284 is a high-density region ($\sim4.1$ stars per arcsec$^2$ as defined in Figure~\ref{fig:completeness_vs_density}). For each region, we show the observed CMD, the corresponding image cutout, and the derived SFH.

From the color images, we can see that each of the regions is visually quite distinct. Region 2284 has a higher density of stars, and the general color is much redder than the other highlighted regions. Region 835 has a clear dust region in the center of the cutout, while Region 6068 has a number of bright blue stars in the two stellar clusters towards the bottom of the cutout. Each of these observations are also present in the CMDs, and SFHs. In contrast, the denser, redder Region 2284, has little star formation in the last 100 Myr, while for Region 835, there is quite a bit of star formation, and a well-populated upper main sequence in the CMD. 


\movetabledown=2.3in
\begin{rotatetable*}
\begin{deluxetable*}{cccccccccccc}
\tabletypesize{\footnotesize}
\tablecaption{Star Formation Histories for PHAST Regions\label{tab:phast_sfh}}
\tablehead{
\colhead{Region ID} &
\colhead{RA$_{\rm BR}$} &
\colhead{Dec$_{\rm BR}$} &
\colhead{Density} &
\colhead{$A_V$} &
\colhead{$dA_V$} &
\colhead{SFR(8.7--8.6)} &
\colhead{SFR(8.3--8.2)} &
\colhead{SFR(7.7--7.6)} &
\colhead{SFR(7.1--7.0)} & 
\colhead{SFR(6.7--6.6)} & 
\colhead{SFR$_{0\text{--}100}$} \\
\colhead{} &
\colhead{(deg)} &
\colhead{(deg)} &
\colhead{(stars arcsec$^{-2}$)} &
\colhead{(mag)} &
\colhead{(mag)} &
\multicolumn{6}{c}{($10^{-5} \times$ M$_\odot$ yr$^{-1}$)}
}
\startdata
7 & 9.887189 & 40.660532 & 0.247 & 0.360 & 0.350 & $4.279^{+0.075}_{-1.919}$ & $1.104^{+0.083}_{-1.104}$ & $0^{+1.197}_{-0}$ & $0^{+3.001}_{-0}$ & $0^{+6.809}_{-0}$ & $0^{+1.568}_{-0}$ \\
8 & 9.887189 & 40.667822 & 0.267 & 0.410 & 0.650 & $5.136^{+1.677}_{-2.560}$ & $0^{+1.492}_{-0}$ & $0^{+1.824}_{-0}$ & $0^{+4.026}_{-0}$ & $0^{+8.880}_{-0}$ & $0.113^{+2.139}_{-0.113}$ \\
9 & 9.887189 & 40.675113 & 0.245 & 0.060 & 0.800 & $0^{+2.134}_{-0}$ & $0^{+0.594}_{-0}$ & $0^{+0.929}_{-0}$ & $0^{+2.744}_{-0}$ & $0^{+6.094}_{-0}$ & $0^{+1.195}_{-0}$ \\
10 & 9.887189 & 40.682404 & 0.270 & 0.110 & 0.750 & $3.035^{+0.126}_{-2.420}$ & $0^{+0.531}_{-0}$ & $0^{+1.077}_{-0}$ & $0^{+2.760}_{-0}$ & $0^{+6.317}_{-0}$ & $0.394^{+1.002}_{-0.370}$ \\
14 & 9.896828 & 40.645950 & 0.296 & 0.160 & 0.800 & $0^{+1.454}_{-0}$ & $0^{+0.860}_{-0}$ & $0^{+1.165}_{-0}$ & $0^{+2.901}_{-0}$ & $0^{+6.670}_{-0}$ & $0^{+1.429}_{-0}$ \\
\enddata
\tablecomments{Excerpt of the machine–readable table of recent star formation histories for individual PHAST regions. 
RA$_{\rm BR}$ and Dec$_{\rm BR}$ give the J2000 coordinates of the bottom–right corner of each region. 
SFR(8.7--8.6), SFR(8.3--8.2), SFR(7.7--7.6), SFR(7.1--7.0) and SFR(6.7--6.6) report the SFR in each logarithmic age bin in units of M$_\odot$ yr$^{-1}$, with asymmetric uncertainties written as $x^{+{\rm err}_{\rm hi}}_{-{\rm err}_{\rm lo}}$. SFR$_{0\text{--}100}$ gives the average SFR over the last 0--100 Myr. 
The full machine–readable table includes all logarithmic age bins from $\log(t/{\rm yr})=8.7$--$6.6$, the corresponding upper and lower uncertainties, and the coordinates of all four corners of each region.}
\end{deluxetable*}
\end{rotatetable*}

\subsection{PHAST SFH Maps} \label{sec:phast_sfh_maps}

We visualize how recent star formation is distributed across the M31 disk by constructing spatially resolved maps of the SFR in multiple time bins. These maps serve as a key tool for identifying the locations, durations, and propagation of recent star-forming activity, 
such as spiral arms, and ring-like structures, that may not be evident from the summed SFHs alone. 

\begin{figure*}
    \centering
    \includegraphics[width=0.99\textwidth]{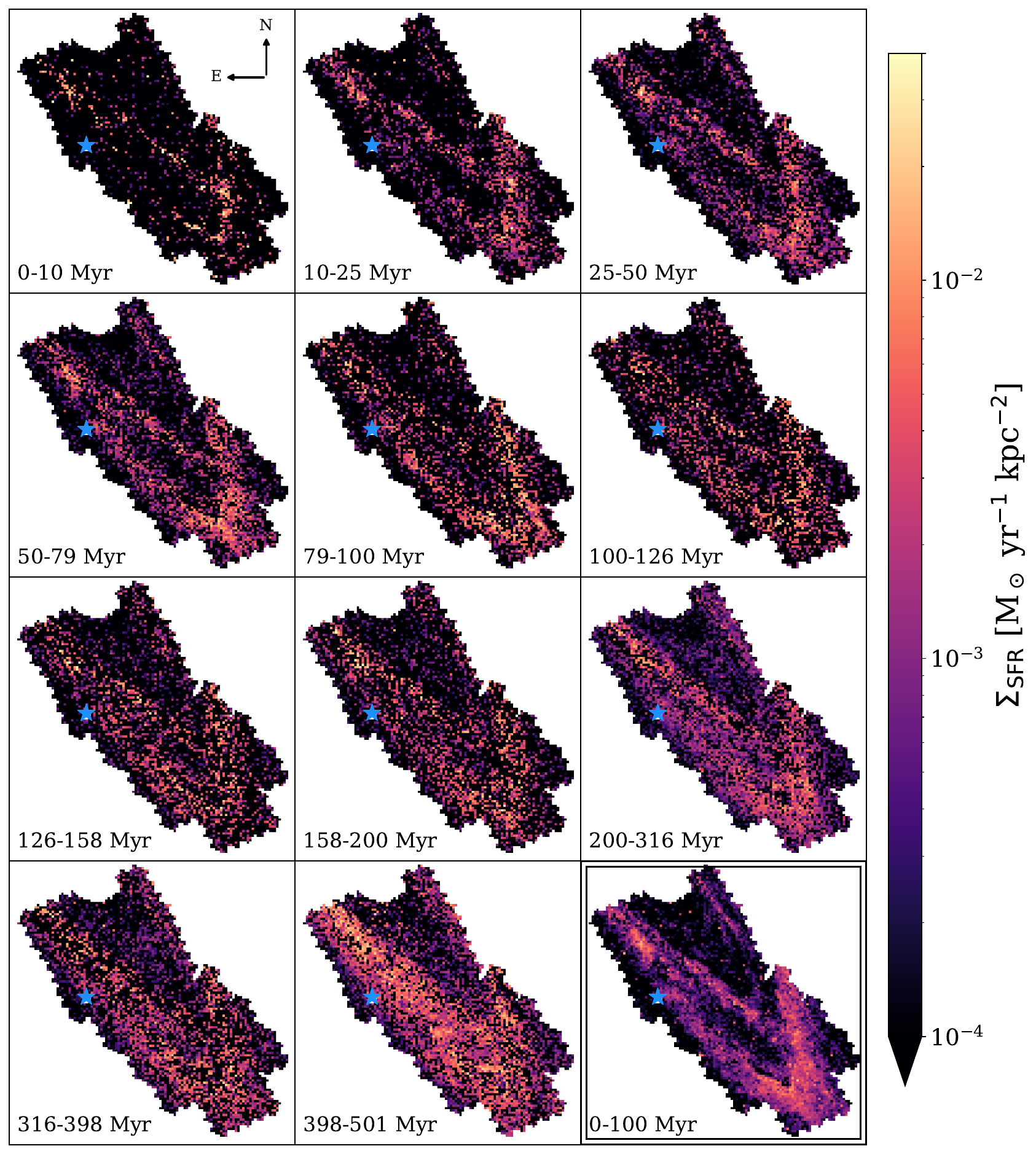}
    \caption{PHAST SFH maps over the indicated timescales in terms of lookback time. The top rows are the most recent epochs, and the subsequent rows look at older lookback times. Each pixels represents one 100 pc$^2$ region, colored by the star formation rate surface density $\Sigma_{\rm SFR}$ in units of M$_\odot$ yr$^{-1}$ kpc$^{-2}$. In each panel, we denote the location of M32 with a blue star. In the bottom right panel, we also include the average SFR over the last 100 Myr, which we refer to throughout the manuscript as $\Sigma{\rm SFR}_{\rm CMD,100}$.}
    \label{fig:phast_maps}
\end{figure*}

Figure~\ref{fig:phast_maps} shows the SFR surface density, $\Sigma_{\rm SFR}$ in each region for a series of time intervals, with the age range labeled in the bottom-left corner of each panel. For each time bin, the region SFR is determined by the total stellar mass formed divided by the time duration, and then normalized by the de-projected area of each region; for the youngest ages ($<100$ Myr), we combine adjacent time bins into broader $\sim$25 Myr intervals to improve signal-to-noise and reveal coherent spatial features. The resulting SFHs are then used to color-code each single spatial region in the given time bin. Regions shown in black have SFRs below $10^{-4}~M_\odot~\mathrm{yr}^{-1}$, a threshold comparable to the typical uncertainty in low-activity regions. As discussed further in Section~\ref{sec: discusion}, the maps reveal strong spatial and temporal variations in star formation across the PHAST footprint. For reference, we also include a map of the SFR integrated over a 100 Myr baseline, which is often used as a canonical timescale for broad-band SFR indicators (bottom right panel). The location of M32 is shown as a blue star in all panels.

\subsection{Stress-testing the PHAST CMD-derived SFHs' Consistency with Observed FUV Fluxes}
\label{sec:synthetic_fuv}

One way to test the validity of the CMD-based SFH methods is to use the derived SFHs to forward-model a synthetic FUV image and compare it directly to the observed \emph{GALEX} map, as was done with great success in \citet{lewis_panchromatic_2017}. In this section, we replicate the methodology of \citet{lewis_panchromatic_2017} to produce a synthetic FUV image, and compare to the \emph{GALEX} map of \citet{martin_galaxy_2005}. To generate synthetic UV fluxes for each PHAST region, we use the \texttt{FSPS} population synthesis code \citep{conroy_propagation_2009, conroy_propagation_2010, conroy_fsps_2010}, following the technique described in detail in Appendix~\ref{appendix_fsps}. 

\begin{figure*}[ht]
    \centering
    \includegraphics[width=0.97\textwidth]{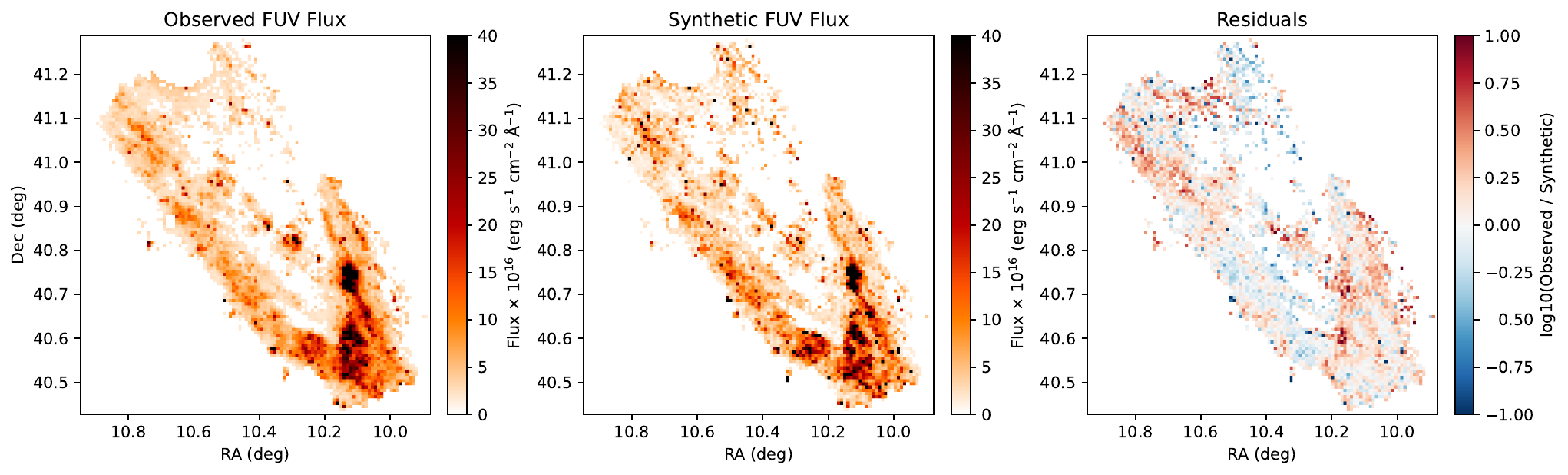}
    \caption{
    Comparison between the observed and synthetic FUV emission in the PHAST footprint. 
    Left: observed \emph{GALEX} FUV image, reprojected to the PHAST spatial grid. 
    Middle: synthetic FUV map constructed by forward-modeling the CMD-derived SFHs with FSPS following the methodology of \citet{lewis_panchromatic_2017}, including a two-component extinction model based on the \citet{cardelli_relationship_1989} law. 
    Right: residual map showing the ratio of observed to synthetic FUV flux. 
    The synthetic map reproduces the major UV-bright structures in the disk and matches the observed FUV fluxes to within a factor of a few over most of the area, indicating that the CMD-based SFHs provide a consistent description of the recent star-forming population. This result is particularly important given the offset in SFR with respect to FUV+24 \micron\ calibrations.
    }
    \label{fig:synthetic_fuv}
\end{figure*}

We present our derived synthetic FUV image, and the comparison, pixel-matched observed \emph{GALEX} image in Figure~\ref{fig:synthetic_fuv}. The left panel displays the observed \emph{GALEX} FUV map, re-gridded to match the PHAST analysis regions. The middle panel shows the synthetic FUV image constructed from the CMD-derived SFHs. The right panel presents the residuals, defined as the ratio of observed to synthetic flux. 

At this level of comparison, the agreement is striking. The synthetic map reproduces the major UV-bright structures, including the 10~kpc ring, the inner arm feature, and many of the smaller-scale star-forming complexes. The residuals are typically within a factor of a few across most of the disk, with larger discrepancies confined to a small subset of regions.
There is no systematic offset in the residuals, which while structured, appear as likely to be high as they are to be low. 

To quantify the degree of agreement between the synthetic and observed FUV maps, we computed the MAD of the squared residuals between the observed and synthetic fluxes. We find this to be 1.8 in our adopted flux units of $10^{16}$ erg s$^{-1}$ cm$^{-2}$ \AA$^{-1}$. The fractional residual distribution, [ (synthetic - observed) / observed ] is centered at -0.11 with a 25th percentile of -0.37 and an 75th percentile of 0.21. Therefore, there is a tendency for the synthetic image to under predict the flux by about 10\%. Although there is scatter, the vast majority of regions agree markedly well within the $\sim$20\% uncertainty associated with UV emission from older stellar populations \citep[e.g.,][]{johnson_measuring_2013}. 

This test provides strong internal validation of the CMD-derived SFHs. Given the combined uncertainties in dust attenuation, UV stellar population modeling, and the CMD-based SFH recovery itself, some level of region-to-region scatter is expected. And yet, despite these uncertainties, the synthetic map successfully reproduces both the observed FUV morphology and the overall flux normalization across the PHAST footprint. This successful replication indicates that the recent SFHs inferred from the CMDs capture the dominant stellar populations responsible for the UV emission.

\subsection{Attaching PHAT SFH Maps to PHAST} \label{sec:phat_phast_sfh_map_attachment}

To place the PHAST results in the broader context of M31’s recent star formation history, we combine the SFH measurements from the PHAT survey \citepalias{lewis_panchromatic_2015} to generate a SFH map that spans the full disk of M31.

\begin{figure*}
    \centering
    \includegraphics[width=0.99\textwidth]{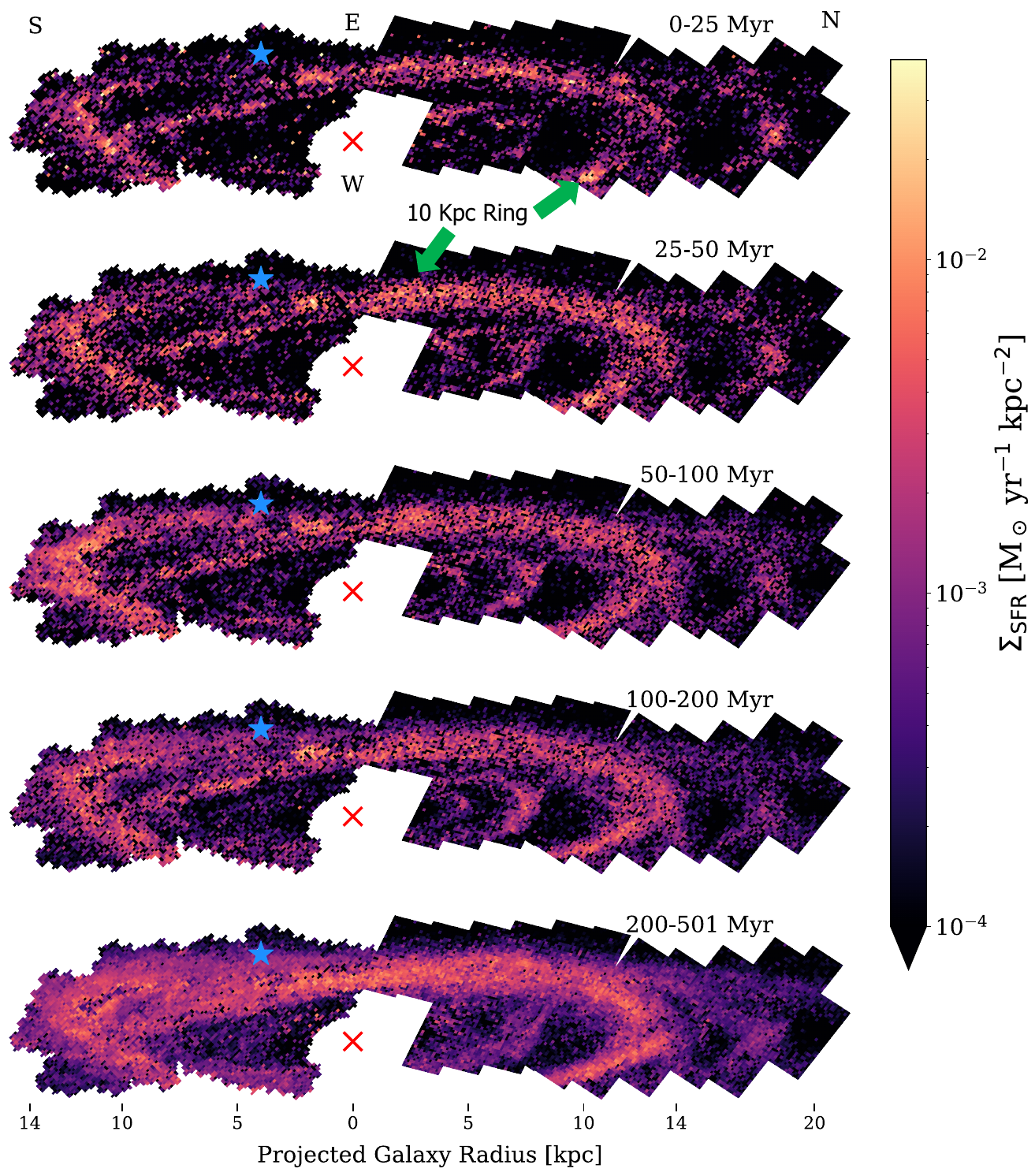}
    \caption{Full M31, PHAT and PHAST combined SFH maps over the indicated timescales. Five time bins are shown moving to older lookback times downward in the plot. The left side of the images is the PHAST region, shown in Figure~\ref{fig:phast_maps}, while the right side is PHAT taken from \citetalias{lewis_panchromatic_2015}. Each pixel is colored by the star formation rate surface density $\Sigma_{\rm SFR}$ in units of M$_\odot$ yr$^{-1}$ kpc$^{-2}$. In each panel, we denote the location of M32 with a blue star, and the center of the galaxy with a red x. Green arrows in the top two panels point to the 10 Kpc ring. An annotation of this Figure, with the time resolution of Figure~\ref{fig:phast_maps} is available in the online version. }
    \label{fig:Full_M31_Map}
\end{figure*}

In Figure~\ref{fig:Full_M31_Map}, we show the spatially resolved map of the SFR surface density, $\Sigma_{\rm SFR}$, for different time bins, combining both the PHAST and PHAT footprints, following Section~\ref{sec:phast_sfh_maps} for calculating the SFR used to color-code the maps. The PHAST data representing the southern half of the disk is shown on the left. The PHAT data in the north was taken from \citealt{lewis_panchromatic_2015}. The position of M32 and the center of the M31 bulge are marked with a blue star and a red "$\times$", respectively.

We note that the original SFRs reported by \citetalias{lewis_panchromatic_2015} do not include the IMF normalization correction described in Section~\ref{sec:imf_limits}. For consistency, we apply the same correction to the PHAT SFHs before combining them with the PHAST results. In practice, we multiply the \citetalias{lewis_panchromatic_2015} SFR in region by the multiplicative factor of 0.779 derived in Section~\ref{sec:imf_limits}. This places both surveys on the same IMF normalization and ensures that any differences seen in the combined maps reflect real spatial variations in the recent SFH rather than a mismatch in the absolute SFR scale. We further note that the PHAT SFHs were derived using an earlier generation of stellar evolution libraries than those adopted here. Specifically, the \citet{girardi_red_2016} transitions for the Padova 2006 isochrones used in this work were not available to \citetalias{lewis_panchromatic_2015}, although the impact on the recent SFHs considered here is expected to be small.

Additionally, the pixel grids for the PHAST and PHAT SFH maps have the same physical pixel size, but are not identically aligned on the sky. The PHAT regions were defined brick-by-brick following the survey layout, whereas the PHAST regions were constructed on a uniform RA–Dec grid and then filtered to retain only cells with full imaging coverage. As a result, the survey boundaries differ in shape, the PHAT survey boundary is straight along the bricks while PHAST appears more jagged. The two footprints overlap only over a handful of pixels from each survey ($\sim10$). In this limited overlap area, the recovered SFHs are consistent, but the pixels are not co-spatial because of the different grid definitions. We therefore stitch the two surveys by plotting the PHAT map directly over the PHAST map in the overlap region, without attempting an additional re-gridding step. Given the very small area involved, any double counting has a negligible effect on the SFRs or spatial trends discussed here. 

The time bins shown in Figure~\ref{fig:Full_M31_Map} are broader than those in Figure~\ref{fig:phast_maps}, and the native age resolution used in the SFH fitting. At that time resolution, the spatial maps appear noisier, with more cell-to-cell variation due to the larger uncertainties in individual bins. Averaging over slightly wider time intervals improves the signal-to-noise and makes the large-scale spatial features easier to identify. We therefore show the maps in these broader age bins to emphasize the coherent structures in the recent SFH. We also note that the 0-100 Myr average is shown in the bottom panel of Figure~\ref{fig:radial}, as opposed to one of the panels in Figure~\ref{fig:Full_M31_Map}. An animated version of the maps at the full native time resolution is available in the online material.

\section{Full M31 Results: Combining PHAT and PHAST}\label{sec:PHAT_PHAST_combined}

The maps shown in Figure~\ref{fig:Full_M31_Map} offer the highest resolution spatially-resolved SFH of M31 measured with resolved stars. Together, these two datasets provide contiguous coverage of two thirds of the star-forming disk, enabling a panoramic view of the past $\sim$500 Myr of star formation in M31.

At a high level, the maps emphasize a number of features that were previously known about M31. First, much of M31's recent star formation is concentrated in the "10 kpc ring" \citep[e.g.,][]{gordon_spitzer_2006, lewis_panchromatic_2015}, which stands out particularly strongly in the north as a coherent, nearly continuous structure traced by active star-forming regions. In the southern PHAST survey region, the ring becomes a more complex mixture of the ring and spiral features, with both a bifurcation and a reduction in strength near M32.

Second, there is a an overall decline in the SFR towards the present, which has been previously noted for the PHAT survey area by \citet{williams_global_2015}. The same effect is seen in the southern PHAST survey area as well. We note, however, that this impression is largely created by the ring, which dominates the bulk of the star formation. We discuss this decline in Section~\ref{sec: global_decline}.

Third, the SFH map highlights large-scale asymmetries in the spatial distribution of star formation across the M31 disk. This asymmetry suggests that different regions of the disk may be responding to distinct dynamical or environmental influences. We explore this further by examining subsets of the disk in Section~\ref{sec:sub_regions}. 

Finally, we note that our decision to match the PHAST SFH calculation to that used previously by \citet{lewis_panchromatic_2015} for PHAT has produced seamless maps despite different observing strategies and depths of the two surveys. In addition, the emergence of coherent structure across thousands of independently-analyzed spatial regions further reinforces the reliability of the measurement technique.

We discuss these above points in more detail below.

\begin{figure*}[ht]
    \centering
    \includegraphics[width=0.95\textwidth]{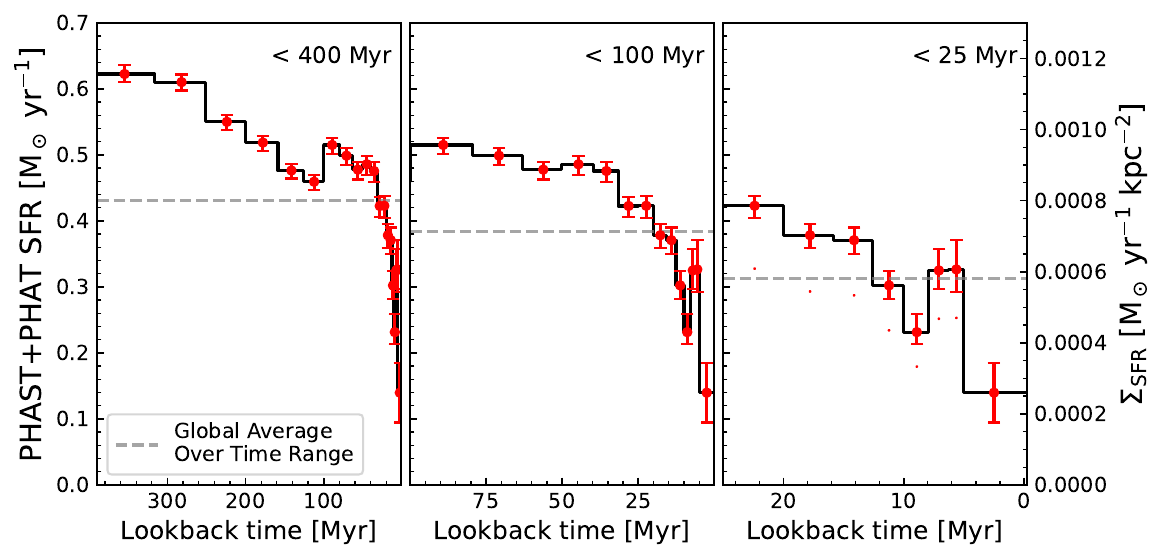}
    \caption{The star formation history of M31 from the PHAT+PHAST as a function of lookback time, where the most recent time is to the right. The black line is the mean SFR per given time bin, while the red lines show the uncertainty for each bin. In each panel, the gray dashed line shows the average SFR over the plotted time baseline. We show both the SFR, and the $\Sigma_{\rm SFR}$ on the right axis. X-axis ranges were chosen to match the complimentary Figure 11 in \citetalias{lewis_panchromatic_2015}. The panels progressively zoom in on the x axis, while the data is the same per bin. We see a pronounced, global decline in the SFR in M31, especially over the last 40 Myr. }
    \label{fig:Full_M31_SFH}
\end{figure*}

\subsection{The Total Recent Star Formation Rate of M31} \label{sec:full_sfh}

We first consider what the combined PHAT$+$PHAST SFHs imply for the total star formation rate of M31. We initially choose to  
integrate our CMD–based SFHs over the last 0–100 Myr, which is taken as a canonical timescale for widely used extragalactic star formation rate indicators such as FUV$+$24$\mu$m \citep[e.g.,][]{kennicutt_dust-corrected_2009}.

Across the union of the two surveys, we measure a total mean SFR of $0.445 \pm 0.006\ M_\odot,{\rm yr^{-1}}$. The uncertainty is derived from adding each regions uncertainty in quadrature, although a larger uncertainty ($\sim 0.03$) could be considered when factoring in additional systematic uncertainty. Scaling the SFR from the combined footprint by the fraction of the galaxy’s FUV D$_{25}$ flux \citep[obtained from ][]{martin_galaxy_2005} enclosed within the footprint implies a total disk–averaged SFR of $\sim0.67\ M_\odot\,{\rm yr^{-1}}$ over the last 100 Myr. The footprint of PHAT+PHAST now covers roughly two–thirds of the actively star–forming disk, so only about one–third of the disk area requires extrapolation.

We note that while this updated extrapolated value for the total M31 SFR is similar to that derived by \citetalias{lewis_panchromatic_2015} based on extrapolating from the PHAT footprint ($\sim 0.7\ M_\odot,\mathrm{yr^{-1}}$) over the same 0–100\,Myr interval, that estimate did not incorporate the IMF correction (Section~\ref{sec:imf_limits}), and is therefore slightly overestimated. Additionally, the star formation intensity is higher in the south (discussed more in Section~\ref{sec:north_v_south}), which would lead an extrapolation based on the north alone to be a slight underestimate. This updated value also relies far less on extrapolation now that a majority of the star–forming disk is directly constrained by CMD–based SFHs. 

These estimates fit comfortably within the range of previous measurements for M31. Other previous studies have examined the global SFR using resolved stars \citep[$\sim 1\ M_\odot,\mathrm{yr^{-1}}$;][]{williams_recent_2003}, 8\,$\micron$ emission \citep[$0.4\ M_\odot,\mathrm{yr^{-1}}$;][]{ barmby_dusty_2006}, FUV emission \citep[$0.6$–$0.7\ M_\odot,\mathrm{yr^{-1}}$;][]{kang_ultraviolet_2009}, H$\alpha$ ($\sim 0.3\ M_\odot,\mathrm{yr^{-1}}$; \citealt{tabatabaei_relating_2010}; $0.44\ M_\odot,\mathrm{yr^{-1}}$; \citealt{azimlu_new_2011}), FUV$+24\ \micron$ \citep[$0.25\ M_\odot,\mathrm{yr^{-1}}$;][]{ford_herschel_2013}, and a combination of several \citep[$\sim 0.4\ M_\odot,\mathrm{yr^{-1}}$;][]{rahmani_star_2016}. 

We note, however, that pinning the ``current'' SFR to the 0-100 Myr average paints an incomplete picture, due to M31's declining recent SFR.
Figure~\ref{fig:Full_M31_SFH} presents the total SFR as a function of lookback time, over all PHAST and PHAT regions. While the individual SFHs of each region exhibit significant diversity -- particularly in the 10 kpc ring and inner disk -- the combined SFH smooths over these fluctuations, revealing robust global trends. We plot this integrated recent SFR on three different timescales to highlight both the longer-term and shorter-term trends. 

In all cases, we find a steady decline in the total SFR over the past $\sim$500 Myr (see also Figure~\ref{fig:older} and discussion in Appendix~\ref{appendix}), the origins of which we will further discuss in Section~\ref{sec: global_decline}. 
This decline makes the integrated SFR a strong function of exactly which time interval one chooses to integrate over. For example, if we average the SFR only over the last 20 Myr, we measure a mean SFR of $0.285 \pm 0.014\ M_\odot,{\rm yr^{-1}}$, which then implies a total disk–averaged SFR of $\sim0.43\ M_\odot\,{\rm yr^{-1}}$ --- roughly $\sim$40\% lower than the 0-100 Myr average. This lower SFR agrees better with the literature measurements based on H$\alpha$ and/or 8 \micron\ emission, which are expected to have shorter associated timescales.

We also note that the existence of an overall declining SFR will change how one interprets quantities averaged over longer baselines like the FUV flux. Such quantities are calculated assuming a constant SFR, which clearly does not apply to M31 as a whole. As such, a FUV based SFR (even corrected with H$\alpha$, or 24 \micron\ flux) will likely underestimate M31's true SFR over the adopted time interval. We make this comparison explicitly in Section~\ref{sec:sfrsd_fuv24} below.

\subsection{Spatial Dependence of Global Trends}\label{sec:spatial}

The recent star formation in M31 is not distributed uniformly across the disk. As shown in Figure~\ref{fig:Full_M31_Map}, the CMD-based SFR maps exhibit strong spatial structure, with star formation concentrated prominently in the ringed-features, and varying across the surveyed area. This is consistent with the PHAT analysis of \citetalias{lewis_panchromatic_2015}, who found that the 10 kpc ring dominated the recent SFH in the northern disk and hosted roughly 60\% of the star formation over the last 400 Myr. The expanded PHAT+PHAST coverage now allows us to revisit this type of analysis over both sides the galaxy and examine how the recent SFH depends on spatial location throughout the disk.

To compare the spatial structure independently of the changing global normalization, we return to the maps in Figure~\ref{fig:Full_M31_Map} and compute, for each age bin, the offset of every region’s SFR surface density from the disk-wide mean at that same timestep. This relative normalization highlights whether a given part of the disk is forming stars above or below the mean at a given epoch, instead of simply reflecting the overall decline in the average SFR. We restrict the comparison to spatial cells with a mean $\Sigma_{\rm SFR}$ over the last 0–500 Myr greater than $6\times10^{-4}\ M_\odot\,{\rm yr^{-1}\,kpc^{-2}}$, so that the normalization is set by regions contributing meaningfully to the recent SFH rather than by the large number of quiescent cells. 

For each time interval, we then calculate $(\Sigma_{\rm SFR}(t)_i-\langle\Sigma_{\rm SFR}(t)\rangle)/\langle\Sigma_{\rm SFR}(t)\rangle$, where $\Sigma_{\rm SFR}(t)_i$ is the SFR surface density in an individual spatial region and $\langle\Sigma_{\rm SFR}(t)\rangle$ is the mean over all cells passing our 0–500 Myr star-forming threshold.

\begin{figure*}[ht]
    \centering
    \includegraphics[width=0.95\textwidth]{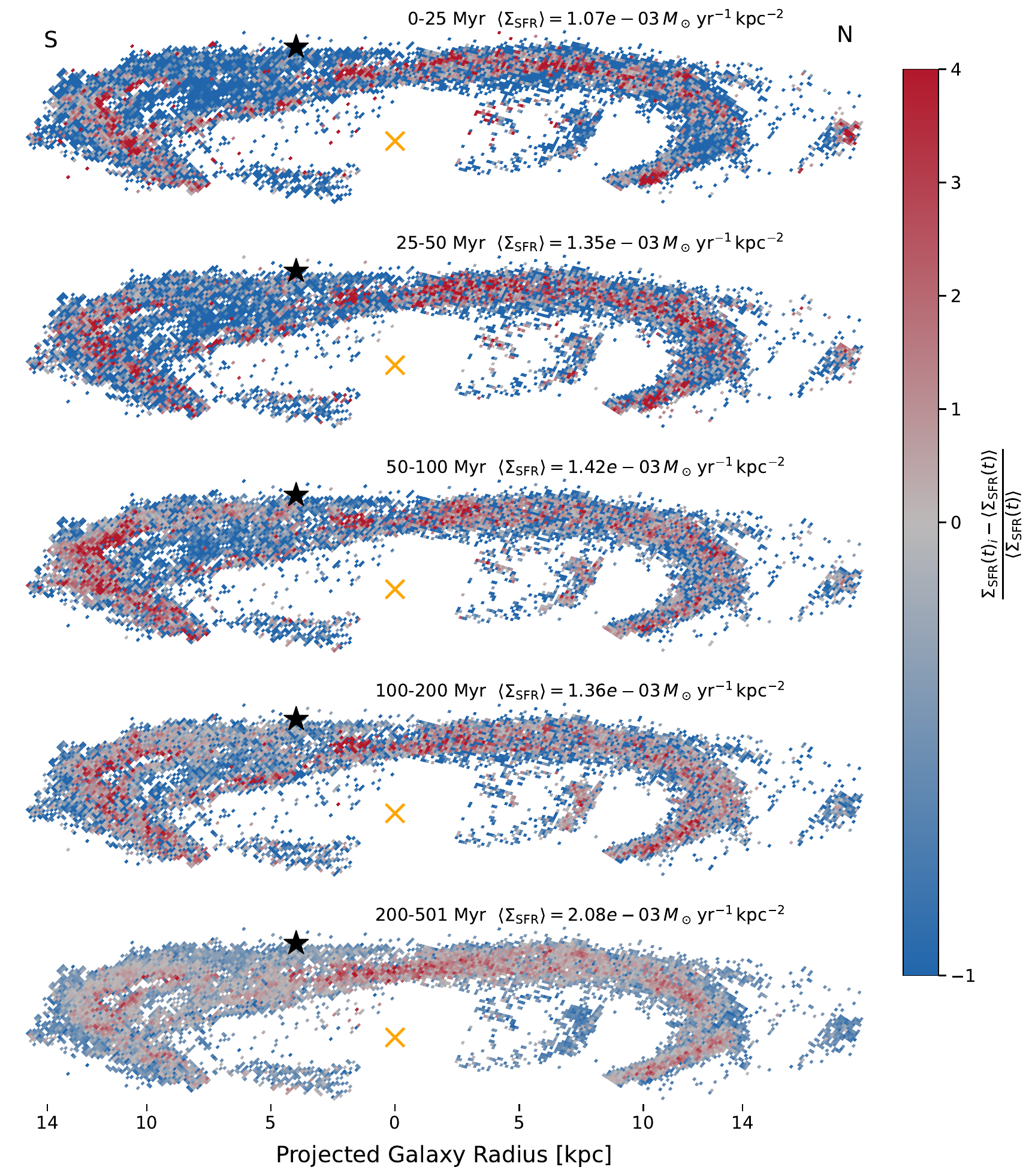}
    \caption{the spatial distribution of recent star formation in M31 as a fractional offset from the disk-wide mean SFR surface density in each age interval. In each panel we plot $(\Sigma_{\rm SFR}(t)_i-\langle\Sigma_{\rm SFR}(t)\rangle)/\langle\Sigma_{\rm SFR}(t)\rangle$, where $\Sigma_{\rm SFR}(t)_i$ is the SFR surface density in an individual spatial cell and $\langle\Sigma_{\rm SFR}(t)\rangle$ is the mean over all cells with $\langle\Sigma_{\rm SFR}\rangle_{0-500\,{\rm Myr}} > 6\times10^{-4}\ M_\odot\,{\rm yr^{-1}\,kpc^{-2}}$. This mask restricts the comparison to regions contributing meaningfully to the recent SFH. Red regions are forming stars above the mean at that epoch, while blue regions are below the mean. The black star marks the center of M32, and the orange cross marks the center of M31. In all age bins, the strongest positive offsets are concentrated in the ring structures, especially the 10 kpc ring. An animated version of this Figure with higher fidelity time steps is available in the online journal. }
    \label{fig:fractional_residuals}
\end{figure*}

The results are shown in Figure~\ref{fig:fractional_residuals}. Gray depicts a region that is forming close to the global average at each timestep, $\langle\Sigma_{\rm SFR}(t)\rangle$, while red regions mark locations forming stars above the mean at a given epoch, and blue regions indicate regions below the mean. 

Even before considering the relative offsets, restricting the sample to regions with $6\times10^{-4}\ M_\odot\,{\rm yr^{-1}\,kpc^{-2}}$ over the last 500 Myr highlights the strongly structured distribution of recent star formation across M31. In the north, the regions with any recent star formation in the last 500 Myr (i.e, included in this plot), are contained almost entirely to the rings, with distinct gaps in between. However, in the south, the regions that have recent star formation are less distinct, and create a much smoother feature. 

This difference in the masked morphology is only part of the picture. Among the regions that remain actively star forming, the strength of the spatial contrast also changes with age. In the oldest time bin, spanning 200–501 Myr, the contrast relative to the disk-wide mean is weaker than at younger ages. However, the map is not spatially uniform, and large-scale asymmetric structure is still clearly present. The persistence of this structure shows that the signal has survived orbital mixing on large scales. Even so, because this broad age bin spans a substantial fraction of the disks dynamical timescale, the present-day positions of these stars are not precise tracers of their birth sites. The most robust interpretation is therefore that the 200–501 Myr map retains information about the large-scale distribution of past star formation, while any finer spatial localization has been blurred by the combination of stellar motions and the wide temporal averaging of the bin.

At younger lookback times, the spatial contrast becomes much stronger and easier to interpret. By 100–200 Myr, and especially 50–100 Myr, the ring structures stand out as the dominant sites of above-average star formation. In the youngest bins, the distribution becomes increasingly heterogeneous, with a smaller number of localized regions, primarily within the rings, accounting for a growing share of the remaining star formation and driving the global average. At the same time, there are some parts of the rings that fall below the global mean, most noticeably around M32, which we will discuss further in Section~\ref{sec:m32_region}. 

The fractional-offset maps make clear that the strongest enhancements in recent star formation remain concentrated in the ring structures over the last $\sim500$ Myr. Having established that relative spatial pattern, we next ask how it translates into the recent SFH budget of the disk, and how the temporal trends that we see in Figure~\ref{fig:Full_M31_SFH} depend on galactic radius. To do so, we analyze the radial dependence of the SFH using radially selected areas on both the northern and southern sides of M31.

We define two matched comparison regions in the northern and southern disk, each with the same projected size, shape, and orientation, and placed along the same axis to sample comparable galactocentric radii on opposite sides of M31. Each region is subdivided into four equal bins, and we then sum the CMD-based SFHs of all spatial cells within each bin. This approach allows a direct comparison of how the recent SFH varies with position through the disk and across the ring structure as a function of galactocentric.

\begin{figure*}
    \centering
    \includegraphics[width=0.95\textwidth]{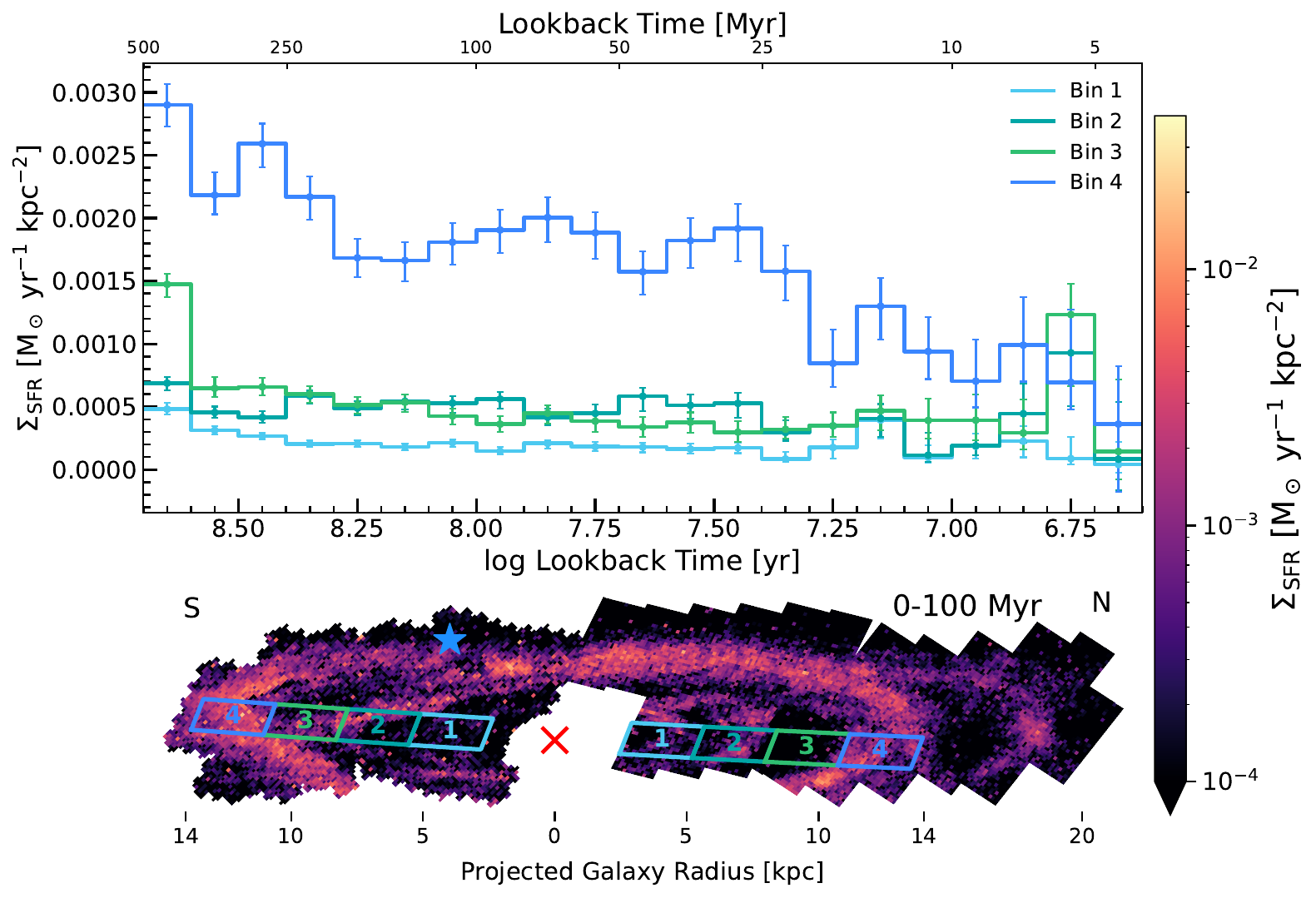}
    \caption{Shown is the radial dependence of the recent star formation history in M31. \textit{Bottom:} map of the mean SFR surface density, $\Sigma_{\rm SFR}$, averaged over the last 100 Myr across the combined PHAT+PHAST footprint. The blue star marks the center of M32, and the red cross marks the center of M31. Two matched comparison regions are overlaid on the northern and southern disk, each subdivided into four bins of equal size. \textit{Top:} SFH each of the four radial bins. In both the north and south, the bins intersecting the 10 kpc ring dominate the recent star formation budget and show the strongest temporal variation, while bins away from the ring remain at substantially lower $\Sigma_{\rm SFR}$, demonstrating that the recent, disk-averaged decline in star formation is driven primarily by declining activity in the ring rather than by a uniform decrease across all radii. }
    \label{fig:radial}
\end{figure*}

Figure~\ref{fig:radial} shows a clear and consistent pattern on both sides of the galaxy. Bin 4, which overlaps the 10 kpc ring structure in the north and the south (bottom panel), coincides with the regions of highest $\Sigma_{\rm SFR}$. As we expect, the top panel shows that bin 4 dominates the recent SFH, with substantially elevated $\Sigma_{\rm SFR}$ relative to the bins interior to the ring. Bin 4 also show the largest temporal variations, where the $\Sigma_{\rm SFR}$ 5 Myr ago is $\sim5\times$ smaller than it was 500 Myr ago. In contrast, bins interior to the ring remain at much lower $\Sigma_{\rm SFR}$ and evolve more gradually.

Therefore, the global recent SFH shown in Figure~\ref{fig:Full_M31_SFH} is strongly shaped by the evolution of rings. Because the majority of the recent star formation occurs in the rings, the disk-averaged decline seen in Figure~\ref{fig:Full_M31_SFH} is driven primarily by declining activity in the ring rather than by a uniform decrease at all radii. The recent wind-down of M31 is therefore best understood not as a featureless fading of the disk, but as a decline within the structures that dominate the star formation budget, and such regions are more spatial concentrated toward the present day.

\subsection{North–South Comparison in the M31 Disk}\label{sec:north_v_south}

While the global behavior across the disk is broadly consistent, with both halves showing the same decline in SFR over the last $\sim 500$ Myr, the spatially resolved maps in Figure~\ref{fig:Full_M31_Map}, and Figure~\ref{fig:fractional_residuals} indicate that the distribution of recent star formation across M31's disk is not entirely symmetric. In fact, PHAST covers key features which are distinctly different from the structures in the northern disk. These include the interface between M31 and M32, the interface between the disk and the giant southern stream, and the split-ring feature in the young populations (See Section~\ref{sec:sub_regions}). M31’s star-forming regions have a more complex morphology in the south, including key features of interest such as the largest star forming region in M31 \citep[NGC~206; e.g.,][]{brinks_ngc_1981}, the most massive young cluster in M31 \citep[e.g.,][]{perina_hstwfpc2_2009}, double the number of bright X-ray sources as the northern half \citep{trudolyubov_discovery_2002, williams_synoptic_2004,vulic_x-rays_2016}, and evidence that velocity dispersion of the southern disk is greater than that measured in the north (D. Grion Filho et al. 2026, in preparation). All of these features confirm the assertion that this half has been more recently or more violently disturbed \citep{gordon_spitzer_2006}. In this section we discuss how these primary differences are probed by their spatially-resolved recent SFHs. 

\begin{figure*}[ht]
    \centering
    \includegraphics[width=0.97\textwidth]{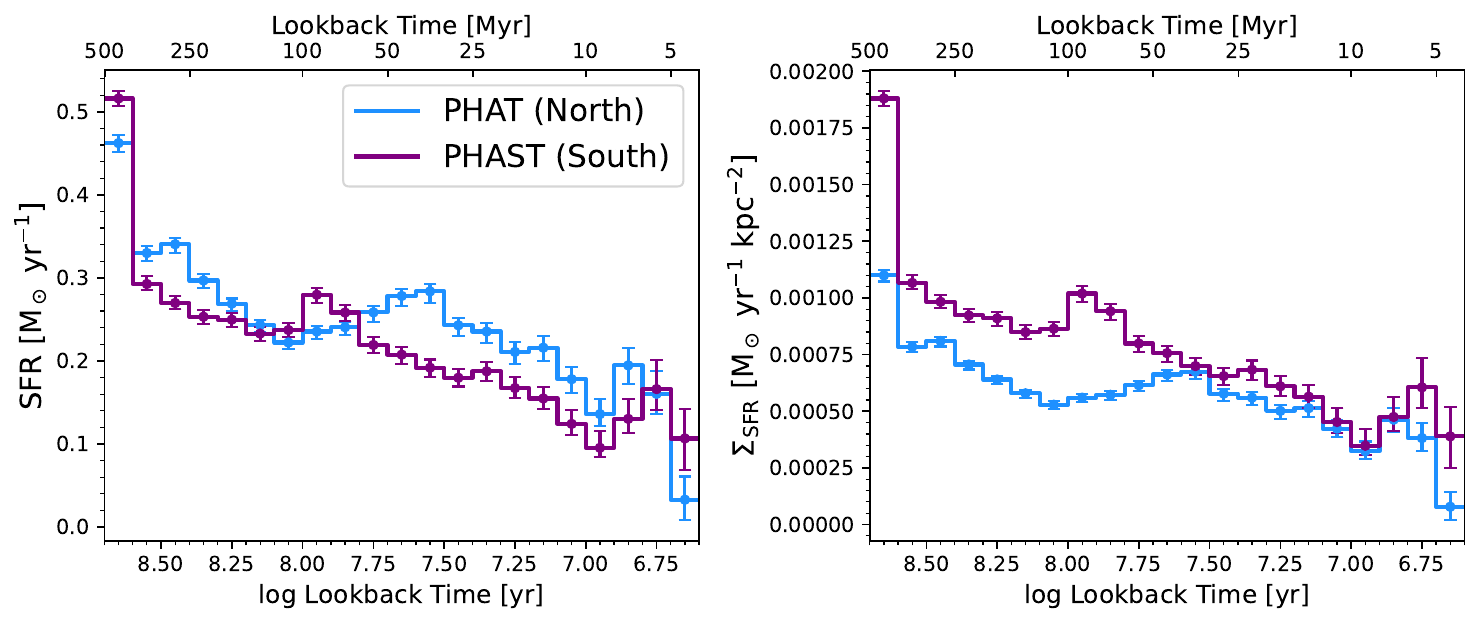}
    \caption{Comparison of the SFHs for the North and South regions of the M31 disk. Left panel shows the SFR, while the right shows the SFR surface density $\Sigma_{\rm SFR}$ as a function of lookback time, where the most recent time is to the right. Blue shows the SFH from the PHAT survey in the North from \citetalias{lewis_panchromatic_2015}, while the purple are the results presented here from the PHAST survey in the South. Both surveys are declining over the 500 Myr window, but the North saw a modest burst from $\sim100-40$Myr ago not seen in the South. We note that measured SFR of $\sim0.5$ M$_\odot$ yr$^{-1}$ in the first bin at log age 8.7 to 8.6 is consistent with the ancient SFH of \citet{williams_phat_2017} derived over the same age interval, and we believe this increase is real (See Appendix~\ref{appendix}).}
    \label{fig:survey_comp}
\end{figure*}

The largest visual discrepancy between the northern and southern maps in Figure~\ref{fig:Full_M31_Map} occurs in the 10kpc ring. In the PHAT footprint, the ring is bright and continuous. While it does continue through the southern disk, it has several distinct gaps in the southern half, most noticeably near M32. However, since the PHAT survey footprint was explicitly designed to follow regions of high UV surface brightness and was truncated where the ring fades, the decline in recent star formation along this boundary is not necessarily surprising. 

To describe the global differences seen in the north and south, Figure~\ref{fig:survey_comp} compares the PHAT and PHAST SFHs directly. The blue curves show the PHAT results from \citetalias{lewis_panchromatic_2015} for the northeastern disk, while the purple curves show the PHAST SFHs presented here for the southern footprint. In the left panel we plot the SFR as a function of lookback time, and in the right panel we show the corresponding SFR surface density, $\Sigma_{\rm SFR}$, averaged over each survey area. The normalization of the curves differ partly because the PHAT footprint covers a larger deprojected area ($\sim 420~{\rm kpc}^2$) than PHAST ($\sim 274~{\rm kpc}^2$), so small differences in $\Sigma_{\rm SFR}$ translate into noticeable changes in the total mass formed.

Over the last 100 Myr, the two halves of the disk form stars at very similar absolute rates. Over this time baseline, we calculate the average SFR in the PHAT survey to be $0.231 \pm 0.004{\rm M_\odot yr^{-1}}$, when considering the IMF correction from Section~\ref{sec:imf_limits}, and adding each regions uncertainty in quadrature. For the PHAST region, we measure a mean SFR of $0.213 \pm 0.004{\rm M_\odot yr^{-1}}$ over the same time interval, despite the smaller surveyed area. This leads to the southern disk having a somewhat higher $\Sigma_{\rm SFR}$, specifically at lookback times of $\sim 80$–100 Myr, as seen in the right panel of Figure~\ref{fig:survey_comp}. This increased $\Sigma_{\rm SFR}$ in the south also agrees with theoretical predictions \citep{hammer_2-3_2018}.

However, over the last $\sim 80$ Myr, the northern and southern disks converge to very similar SFR surface densities. The most noticeable difference in Figure \ref{fig:survey_comp} is that the PHAT region has a steady increase in star formation from 100 Myr to 40 Myr ago, from $\sim0.22$ M$_\odot$ yr$^{-1}$ to $\sim0.29$ M$_\odot$ yr$^{-1}$, that has no clear analogue in the PHAST measurements. In contrast, the southern disk instead shows a smoother decline in the SFR from 100Myr to the present.

Within the uncertainties, the PHAT and PHAST SFHs at $\lesssim 40$~Myr are consistent with a shared picture in which the M31 disk has experienced a recent decline in star formation, superimposed on smaller-scale spatial variations. At the youngest ages however, both surveys show a modest upturn in SFR in the last $\sim10$ Myr relative to the 10–20 Myr bins. This feature is intriguing, but should be interpreted with caution, since this is the regime with the highest systematic uncertainties (Section~\ref{sec: uncertainty}). 

We also note that the SFR in the oldest bin of our recent-SFH measurement may appear unusually high, roughly 60\% higher relative to the bins immediately below it. However, we believe this is a real feature in the M31 SFH. The measured SFR in the log-age 8.7–8.6 bin agrees with the ancient SFH of \citet{williams_phat_2017} over the same interval, providing an important consistency check at the boundary between the two analyses. As discussed further in Appendix~\ref{appendix} and Section~\ref{sec: global_decline}, the SFR at this time is also consistent with the SFH recovered in the regions where we can reliably probe to older ages, supporting the interpretation of a longer decline in M31’s SFH.

\section{Insights into M31's Evolution}\label{sec: discusion}
The SFHs derived from the PHAST and PHAT surveys provide a coherent view of how recent star formation has unfolded across the M31 disk and open several avenues where these maps can be used as diagnostic tools. In this section we use the M31 SFH maps combining the PHAT and PHAST regions to discuss how the recent SFR has evolved globally, and to discuss possible mechanisms responsible for the observed decline. In addition, we highlight two intriguing subregions in the PHAST footprint: the M32 region and the inner arm structure, as case studies of how local environment shapes the SFH.

\subsection{Assessing the Global Decline in the SFR} \label{sec: global_decline}
As shown in Figure~\ref{fig:Full_M31_SFH}, the full SFH for the PHAT+PHAST combined footprint shows a very distinct picture: M31 has experienced a clear decrease in SFR over the last 500 Myr, with a particularly steep drop in the last $\sim 40$~Myr. When we integrate over the full PHAT+PHAST footprint, the average SFR declines from $\sim 1\,{\rm M_\odot\, yr^{-1}}$ at lookback times of $\sim 500$~Myr to $\simeq 0.5{\rm M_\odot\, yr^{-1}}$ 40 Myr ago. 

The orbital timescale of M31’s star–forming disk is of order $\sim250$ Myr, so a coherent decline in SFR across most of the disk over only $\sim40$ Myr would appear, at face value, uncomfortably rapid. This makes it particularly important to ask whether part of the apparent decline could arise from systematics in the very youngest age bins, where stellar evolution models, dust treatment, and crowding effects introduce the largest uncertainties. However, multiple lines of evidence indicate that the decline is unlikely to be a methodological artifact, and is instead physical.

First, the downward trend is already present when we average the SFH over 40–500 Myr, well beyond the regime where massive–star evolutionary uncertainties and binary physics are most acute \citep[e.g.,][]{dolphin_numerical_2002,gallart_adequacy_2005,dolphin_estimation_2013} and where we see the largest mismatch between evolutionary models (Section~\ref{sec: uncertainty}). Second, the same \texttt{MATCH}-based fitting pipeline applied to M33 in the PHATTER survey recovers a recent SFH that is comparatively steady over the last 100 Myr and does not show an analogous global decline, even though the CMD depth, modeling assumptions, and extinction treatment are very similar \citep{lazzarini_panchromatic_2022}. This argues that the behavior we see in M31 is not a generic bias in the technique, but rather a feature of the galaxy. Finally, our extinction maps and CMDs do not show any abrupt change in dust geometry or fit quality at the lookback times where the SFR begins to drop, which would be expected if the decline were primarily driven by a breakdown in the dust modeling. For these reasons, we believe the observed decline is a real feature of M31's SFH history, and in this section, we discuss possible explanations.

\subsubsection{Is M31 Running out of Gas?}
From a gas-supply standpoint, a gradual decrease in SFR over several hundred Myr is compatible with the timescale of molecular gas depletion, of order 1–3 Gyr, seen in nearby star-forming disks \citep[e.g.,][]{bigiel_constant_2011,schruba_molecular_2011, sun_star_2023}. However, it is important not to interpret this trend as a literal ``running out of gas’’. In both observations and simulations, galaxies like M31 do not exhaust their fuel reservoir in isolation unless an additional process suppresses accretion and or prevents recycled material from re-cooling efficiently \citep[e.g.,][]{bell_galaxy_2008, somerville_physical_2015}. Empirically, truly quenched bulge-less central galaxies are rare, which suggests that sustained suppression of star formation is linked to compact central structures and the physics associated with them \citep{bell_galaxy_2008}.

This causality is explicit in modern galaxy-formation models where without super massive black-hole (SMBH) feedback, simulations generally fail to quench massive central galaxies and instead retain excessive late-time star formation \citep[e.g.,][]{ byrne_effects_2024}. In IllustrisTNG, for example, SMBH feedback is explicitly responsible for quenching galaxies in intermediate and high-mass halos \citep{pillepich_simulating_2018}, and across multiple simulation suites the most predictive parameter separating star-forming from quenched central galaxies is the SMBH mass, consistent with cumulative, integrated feedback \citep{piotrowska_quenching_2022}. Observationally, direct SMBH mass measurements similarly show that quiescence correlates strongly with $M_{\rm BH}$ at fixed stellar mass \citep{terrazas_quiescence_2016,terrazas_supermassive_2017}.


For M31, the central SMBH (M31$^*$) has $M_{\rm BH}\simeq 1.4\times10^{8}\,M_\odot$ \citep{Bender2005}, but the nucleus today is extremely low luminosity \citep{li_murmur_2011}, so any direct, contemporaneous feedback on the disk is expected to be weak \citep{Li2009}. Constraints on past activity instead come from relic diagnostics. XMM–Newton/RGS spectroscopy of M31's inner bulge has been interpreted as evidence for an AGN episode as recently as roughly 0.5 Myr ago \citep{Zhang2019}, and extended high-energy emission has been discussed as potentially consistent with past nuclear injection, though interpretations are not unique \citep{Pshirkov2016}. In addition, deep very long baseline interferometer (VLBI) observations indicate that the nuclear radio emission is intrinsically extended on milliarcsecond scales, consistent with a hot outflow/wind from the weakly accreting AGN \citep{Peng2023}. This evidence is qualitatively consistent with ``maintenance-mode’’ feedback, in which energy and momentum injection keep the halo and inner-disk gas hot, turbulent, or otherwise unable to cool and condense efficiently \citep[e.g.,][]{Harrison_review, Alexander_review}. In such instances, star formation can decline without wholesale expulsion of the star-forming ISM, because the long-term resupply of cold gas is reduced and recycling times are lengthened. 

These considerations motivate AGN-related maintenance feedback as a physically plausible contributor to M31’s long-term decline in SFR and central quiescence. However, establishing a direct causal link between AGN activity and a galaxy’s star-formation history is intrinsically difficult due to spatial and timescale mismatches \citep[e.g.,][]{Alexander_review, leung_spatially_2026}. 

\subsubsection{Late Stages of a Longer Decline}

A decline of a factor of a few over the last few hundred Myr is also consistent with what is expected for green valley galaxies like M31. Empirical reconstructions and large–survey analyses generally find that the post star-burst ``wind-down'' is not instantaneous, but proceeds over a range of characteristic timescales from several $\times 10^8$ yr to a few Gyr, with galaxy–to–galaxy diversity \citep[e.g.,][]{martin_uv-optical_2007,schawinski_green_2014, lin_almaquest_2026}. 

When we place our measurements of the recent SFH in the broader context of M31's evolution, the recent decline appears to be the continuation of a longer–term evolution of the M31 disk. The ancient SFH derived from the PHAT survey shows that M31 experienced a strong enhancement in star formation $\sim2$ Gyr ago, during which the global SFR reached upwards of $6 {\rm M_\odot,yr^{-1}}$, followed by a gradual decrease toward quiescence \citep{williams_global_2015,williams_phat_2017}. In fact, when we consider the outermost regions of PHAST where we can trace out to ages of 1.2 Gyr with the main sequence turn-off (See Appendix~\ref{appendix}), we see a fairly constant decline in the SFR in the last 1.2 Gyr. 

Our measurements here extend this picture to younger ages, showing that the disk has continued to evolve toward lower SFRs over the last 500 Myr, with a modest enhancement around 100 Myr, followed by the more pronounced decline in the last $\sim 40$~Myr. In this view, the current low SFR is not the result of a sudden, synchronized shutdown across the disk, but rather the latest stage of a multi–Gyr relaxation from a previously more active phase.

In fact, the decline is not perfectly uniform across the disk. As discussed in Section~\ref{sec:north_v_south}, and seen in Figure~\ref{fig:survey_comp}, the PHAT region sees a modest increase in star formation from 100 Myr up until 40 Myr ago, which is not seen in the PHAST survey. More importantly, inspection of both Figure~\ref{fig:Full_M31_Map} and the radial analysis in Figure~\ref{fig:radial} show that most of the recent star formation is concentrated in the rings, which dominates the total star formation budget for the galaxy. The disk-averaged decline in Figure~\ref{fig:Full_M31_SFH} is therefore driven primarily by declining activity in the ring, rather than by a uniform decrease across the disk. 

For the reduction in the 10 kpc ring, a plausible interpretation is that the recent decline reflects reduced fueling. In dynamical models, long-lived rings persist when gas replenishment balances inward transport and depletion by star formation, so a reduction in inflow can lower the SFR while leaving the ring morphology intact \citep[e.g.,][]{seo_star_2013, dekel_origin_2020}. This same balance also helps set the lifetime of the structure. Star-forming rings are generally expected to be transient, but may persist up to order-Gyr timescales when maintained by the underlying disk dynamics \citep[e.g.,][]{buta_galactic_1996,sellwood_lifetimes_2011,sellwood_spirals_2022}. 

More broadly, M31 offers a stringent test for galaxy-evolution models. Any successful framework must reproduce not only the present-day global SFR, but the structure and longevity of the rings, the quiescent inner regions, and the recent decline over the last couple of Gyr. As a result, M31 serves as a key benchmark for simulations in evaluating how processes such as gas accretion, recycling, and maintenance-style AGN feedback combine to regulate star formation in massive disk galaxies transitioning through the green valley.

\subsection{Specific Regions of Interest in the PHAST Survey}\label{sec:sub_regions}

The complete SFH map for PHAST presented in Figure~\ref{fig:phast_maps} show specific areas of the southern disk that stand out in contrast to the consistently structured north. Sub-selecting these more localized features and comparing the SFH of those regions to that of the global average offers additional insight into the underlying physical drivers of the SFR. In this section, we highlight two distinct sub-regions within the PHAST footprint: the region surrounding M32, and a segment of the inner-arm region of the 10 kpc ring split where, in our projection space, there appears to be a connecting spur between the inner and outer rings. These regions were chosen for their distinct structural and environmental context, and for their contrasting recent SFH behavior. By examining them in more detail, we aim to link small-scale structure to the broader trends seen across the southern disk and, in the case of M32, to place its recent activity in the context of the satellite’s interaction history with M31.

\begin{figure*}
    \centering
    \includegraphics[width=0.95\textwidth]{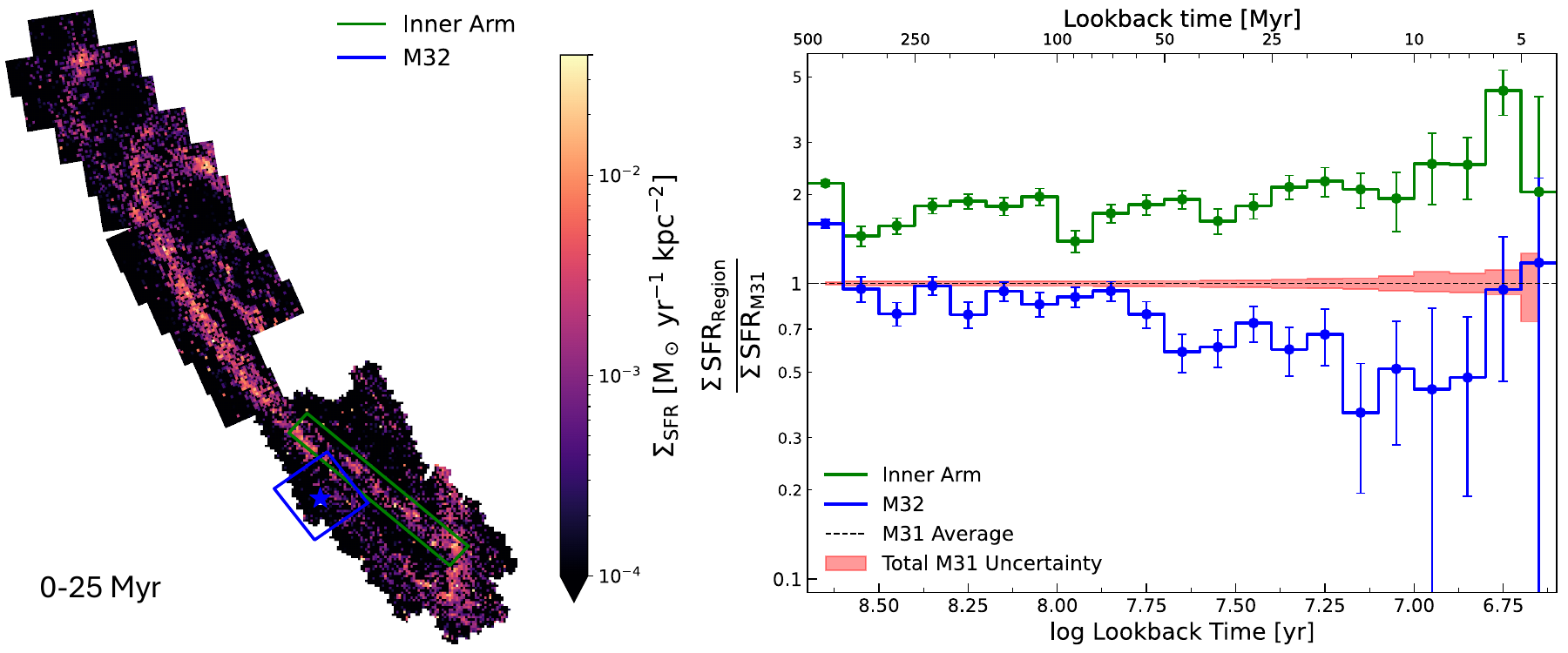}
    \caption{Subregions used to investigate local variations in the recent SFH. 
    \emph{Left:} SFR surface density map for the 0–25 Myr time bin from Figure~\ref{fig:Full_M31_Map}, with the two subregions highlighted. The blue outline marks the M32 region and the green outline marks the inner-arm/bar region near the apparent spur connecting the inner and outer 10 kpc rings. 
    \emph{Right :} Ratio of the SFR in each subregion to the global SFR averaged over the combined PHAT+PHAST footprint as a function of lookback time. The horizontal dashed black line indicates parity with the global average; the red shaded band shows the uncertainty on the global SFH from Figure~\ref{fig:Full_M31_SFH}. The curves for the M32 and inner-arm regions illustrate how their recent SFHs diverge from the disk average, highlighting a recent suppression of star formation in the M32 region and distinct burst in the inner arm.
    }
    \label{fig:subregion_comp}
\end{figure*}

These regions are shown in the left-hand panel of Figure~\ref{fig:subregion_comp}, where we outline the selected areas on top of the 0–25 Myr SFR surface density map from Figure~\ref{fig:Full_M31_Map}. The M32 region is outlined in blue and the inner-arm region in green, while M32 itself is marked with a blue star. While the exact boundaries of these regions were defined somewhat arbitrarily, we experimented with modestly larger and smaller apertures and found that the quantitative SFH behavior is stable within the uncertainties. 

\subsubsection{The M32 Region}
\label{sec:m32_region}

The first area of interest, and arguably the most striking structure in the southern part of M31, is the region surrounding the compact satellite M32. In Figure~\ref{fig:subregion_comp}, this subregion is outlined in blue. The aperture shape was chosen to encompass the area immediately surrounding M32 and the associated perturbation in the disk.

In the right hand panel of Figure~\ref{fig:subregion_comp}, we compare the SFR within this region to the overall SFR across the PHAT and PHAST surveys as a function of lookback time. We take the ratio such that 1 is the galactic average, which is shown by the dashed black line. The red shaded region around this line is the uncertainty in this measurement, using the same error-bars from Figure~\ref{fig:Full_M31_SFH}.

From $\sim 500$~Myr to $\sim 60$~Myr ago, the M32 region is statistically consistent with the M31 average, with only mild fluctuations around unity. At younger ages, however, the behavior changes. Starting at $\sim 60$~Myr, the SFR in this region drops significantly below the global average and remains suppressed. From $\sim 50$~Myr ago, to $\sim 20$~Myr ago, the region around M32 is quenched by $\sim 40\%$ compared to the rest of the galaxy, reaching a minimum of roughly $65\%$ of the mean disk SFR at a lookback time of $\sim 15$~Myr. This pattern suggests a recent and localized reduction in star formation activity that is not mirrored across the galaxy as a whole. In fact, in Figure~\ref{fig:fractional_residuals}, this regions stands out as particularly quenched with respect to the rest of the 10 kpc ring.

The seemingly most plausible explanation for this localized quenching would be a response to an interaction with M32. Several studies have linked M32 to M31’s ring-like structure and disturbed recent SFH, but the interaction history remains uncertain \citep[e.g.,][]{gordon_spitzer_2006,block06,dierickx14}. A long-standing limitation has been the lack of a well-constrained proper motion and an accurate line-of-sight distance for M32, leaving substantial freedom in orbital reconstructions. As a result, published models span a range of encounter geometries and epochs, including scenarios with a disk passage $\sim 200$ Myr ago \citep{block06} and alternatives that place the most recent strong interaction much more recently \citep{gordon_spitzer_2006}.

Recent HST data have been obtained through GO-15658 (PI: S.T. Sohn) to directly address this issue, and measure the proper motion of M32 relative to that of M31's for the first time. Preliminary evidence suggests that M32 is moving southward relative to M31 with significant velocity (Fardal et al. 2026, in prep.). Additionally, ongoing work to integrate the backwards orbit of M32 in a rigid M31 halo potential suggests that a passage through M31's disk was possible $\sim20$ Myr ago (with uncertainties of $\sim10$ Myr). However, further analysis is needed to determine the effects of tidal stripping of M32, and the range of orbital uncertainties derived from considering a range of relative distances between M32 and M31 (Patel et al. 2026, in prep.). These proper motion results (Fardal et al. 2026, in prep), are consistent with the earlier model of \citet{gordon_spitzer_2006}, which suggest M32's passage through M31's disk occurred 10~Myr ago. 

If the localized quenching seen in Figure~\ref{fig:subregion_comp} is driven by an interaction with M32, then the timing in our resolved SFH is more aligned with a relatively recent perturbation than with a passage $\sim 200$ Myr ago. In particular, the lowest measured SFR in this region occurs at lookback times of order $\sim 15$ Myr, and we measure a modest uptick in the last $\sim 10$ Myr, qualitatively consistent with the timescales invoked in recent-interaction scenarios. At the same time, our uncertainties in these youngest age bins are the largest, and the measured SFRs remain formally consistent with continued quenching within the errors. 

However, the SFH results presented here do not provide clear support for the very recent interaction scenario either. Importantly, the SFR suppression in this region begins earlier, at $\sim 60$ Myr lookback time, before the epoch implied by the most recent-interaction models. This timing either implies that M32's projected distance from the disk is an underestimate and it is actually further away, or a simple one-event explanation doesn't capture the full SFH of this region. The early decline may reflect local evolution unrelated to M32, or it may indicate that multiple processes act together, with any M32-driven perturbation superposed on pre-existing changes in the local star-forming environment.

In either case, these resolved SFH constraints presented here should play a central role in future modeling of the M31–M32 system. A full hydrodynamical treatment that predicts the timing, amplitude, and spatial extent of the disk’s star-formation response to an M32 passage would provide a critical test to the degree to which an interaction could account for the quenching history we measure in this region.

\subsubsection{The Inner Arm Region}
\label{sec:inner_arm_region}

In contrast to the quenched behavior of the M32 region shown above, the inner arm ring segment describes a different picture, despite only being a short distance away in projection from the M32 field. In Figure~\ref{fig:subregion_comp}, the inner arm region is outlined in green, tracing a segment of the inner spiral structure where the projected morphology suggests a narrow bar that links the inner and outer portions of the 10~kpc ring.

Over the last $\sim 500$ Myr, the inner-arm region sits systematically above the global average. This enhancement is consistent with the expectation that spiral arms and ring-like structures tend to host elevated star formation relative to the surrounding disk \citep[e.g.,][]{kennicutt_star_1989, kennicutt_star_1998, querejeta_spiral_2024}. Notably however, this feature is not simply part of the main 10 kpc ring, but a distinct segment of the split southern morphology that sustains enhanced star formation for several hundred Myr. That persistence is significant because it extends beyond an orbital timescale, implying that the feature has avoided being erased by radial and azimuthal mixing over much of that interval. This inner-arm component therefore traces a genuine, long-lived dynamical structure in the southern disk, rather than a short-lived fluctuation or a purely projected connection between neighboring star-forming regions.

Within this long-lived ring structure, the inner-arm segment also shows a pronounced enhancement at the youngest ages.\footnote{We note that the size of the segment shown here was chosen to try and avoid ambiguity in determining which exact features are a part of the inner-arm. Although the region we adopt includes the star-forming complex NGC~206 at its south–western edge, we tested smaller apertures focused solely on the inner–arm structure and found similar recent SFR enhancements within the uncertainties. The more spatial regions included increase the signal to noise and are therefore easier to interpret.} This behavior stands in sharp contrast to the neighboring M32 region, where the SFR in the last 60 Myrs is suppressed relative to the disk average (Section~\ref{sec:m32_region}). Additionally, the youngest ($< 50$ Myr) panels of Figure~\ref{fig:radial} show the inner-arm segment appears to be forming stars more actively than the nearby portion of the 10 kpc ring at similar projected galactocentric radius. The juxtaposition of an actively star-forming inner arm and a locally quenched M32 region within a few kpc highlights the sensitivity of recent star formation to local dynamical and environmental conditions, even inside a long-lived ring structure.

Interestingly, the apparent simplicity of the inner arm in projection does not reflect the true three-dimensional structure of the gas and stars. In our SFH maps, this feature resembles a straight bar connecting the inner and outer rings in the PHAST footprint, and clearly seen by the massive dust lane in Figure~\ref{fig:showing_diff_regions}. However, the underlying kinematics indicate a more complex configuration. The H\,\textsc{i} velocity structure in the vicinity of this inner-arm segment from the new VLA data from LGLBS \citep{koch_karl_2025} shows possible evidence for a coherent set of velocities that are distinct from the neighboring ring material, similar to the behavior identified in earlier H\,\textsc{i} studies of M31 \citep[e.g.,][]{braun_distribution_1991}. This pattern suggests that the inner arm is not simply a bridge between two rings. Instead, it likely traces a geometrically distinct ring or arm segment that is inclined relative to the bulk of the disk and projected into an apparently straight feature on the sky.

These results paint a picture in which two nearby regions in projection space, may be more separated than initial indications suggest. The M32 region shows a significant decline in SFR over the past $\sim 60$~Myr, while the inner arm region exhibits a persistent enhancement during the same period relative to the galaxy wide average. Importantly, the quenching mechanism of the M32 region, did not impact the inner arm region. 

The fact that the inner arm appears to be a distinct dynamical structure, rather than a mere continuation of the outer ring, makes this contrast more illuminating. This SFH result point out that these seemingly nearby regions experienced very different recent evolutionary histories. This combination of elevated SFR and unusual geometry marks the inner arm as a particularly interesting site for probing how local dynamical structures and external perturbations shape the recent SFH of M31, and should be the focus of future studies.


\section{Comparison to Other Star Formation Tracers}
\label{sec:other_tracers}

The CMD-derived SFHs provide a view of star formation that is conceptually different from that obtained with traditional integrated luminosity tracers, particularly those with additional empirical corrections for dust, such as FUV+24 \micron\@. In contrast to the CMD-based SFHs derived here, which reconstruct the distribution of stellar ages and masses over a grid of lookback times, while other tracers infer instantaneous SFRs from the luminosity of short-lived massive stars and dust re-radiation, implicitly assuming a steady SFH and a fixed dust geometry \citep[e.g.,][]{kennicutt_dust-corrected_2009,boquien_towards_2016}. 

Several recent studies have highlighted a systematic offset between these approaches, with FUV based SFR indicators yielding results that are smaller by factors ranging from 1.3 to 3 compared to those inferred from resolved stellar populations \citep[e.g.,][]{mcquinn_calibrating_2015,lewis_panchromatic_2017, lazzarini_panchromatic_2022} \footnote{We note that the \citet{lazzarini_panchromatic_2022} study did not include the finite IMF integration limits, and if we adopt the same IMF limits chosen here, the SFRs would be interpreted $\sim$ 20\% lower than the published values. When taking into consideration this correction, the CMD based SFRs are brought closer into alignment with the FUV based SFR, but the CMD estimates are still a factor of 1.3 larger.}. Since both methods are widely used and each rests on well-motivated assumptions about stellar populations, this mismatch has large impacts for the community. 

In this section, we first compare FUV+24 \micron\ SFR surface densities to the CMD-based SFRs averaged over the last 100 Myr. We then test whether either FUV+24 \micron\ or CMD derived SFR values held constant over the last 100 Myr can produce a synthetic FUV image similar to the observed \emph{GALEX} FUV emission. 

\subsection{CMD-based SFR Surface Density Compared to UV Luminosity Tracers}
\label{sec:sfrsd_fuv24}

Star formation rates can be estimated in many different ways, including H$\alpha$ emission, UV continuum, infrared emission, combined UV+IR recipes, or non-thermal synchrotron radio emission \citep[e.g.,][]{kennicutt_star_1998, kennicutt_star_2012, calzetti_star_2013}. The most direct external comparison to our CMD-based recent SFHs are the tracers that involve UV emission, as both are grounded in similarly young stellar populations \citep[e.g.,][]{kennicutt_star_2012}. In contrast, H$\alpha$ emission traces the ionized-gas response to the youngest, most massive stars, and is subject to much greater stochasticity and dependence on the properties of the ambient ISM. 

We therefore compare our CMD-based SFHs to the FUV+24 \micron\ calibration from \citet{2008AJ....136.2782L},

\begin{equation} \label{eq:fuv_24mic}
\Sigma{\rm SFR}_{\rm{FUV}+24 \mu m} = \Bigl( 8.1\times10^{-2}\, I_{\rm FUV}
                       + 3.2^{+1.2}_{-0.7}\times10^{-3}\, I_{24} \Bigr) \cos i,
\end{equation}
where $I_{\rm FUV}$ and $I_{24}$ are the FUV and 24 \micron\ intensities in MJy\,sr$^{-1}$. The cos($i$) term takes into account M31's inclination $i$, taken to be $77\degree$ \citep{walterbos_optical_1988}. $\Sigma{\rm SFR}_{\rm{FUV}+24 \mu m}$ is therefore a SFR surface density with units of M$_\odot$ yr$^{-1}$ kpc$^{-2}$, directly comparable to the CMD analysis from Section~\ref{sec:methods}. We calculate $\Sigma{\rm SFR}_{\rm{FUV}+24 \mu m}$ using the 24 \micron\ map from the \emph{Spitzer}/MIPS observations of \citet{gordon_spitzer_2006}, and the FUV map from the \emph{GALEX} imaging of \citet{martin_galaxy_2005}. We project both data sets onto the PHAST spatial grid by integrating the native-image flux over each PHAST region, weighting each pixel by the fraction of its area that falls inside the region, making sure flux is conserved. We then compute $\Sigma{\rm SFR}_{\rm FUV+24  \mu m}$ in each spatial bin.

We note two adjustments to the \emph{GALEX} image, needed to derive $I_{\rm FUV}$ in a way consistent with the \citet{2008AJ....136.2782L} study. First, we subtract a uniform background from the image based on a section of non-galaxy pixels, and second, we correct for foreground dust. We choose to correct for dust based on each region's best fit foreground $A_V$ (Figure~\ref{fig:av_dav}) from the CMD fits. Alternatively, we could have chosen a uniform dust correction based on the \citet{schlegel_maps_1998} maps of nearby regions, rescaling them following \citet{schlafly_measuring_2011}; however, the region-by-region correction produced the best comparison to our CMD analysis. 

For the CMD-based reference SFRs, in each region, we integrate the SFHs over the last 100~Myr to derive the mean SFR (Section~\ref{sec:phast_sfh_maps}). This timescale is chosen to match what is generally considered the characteristic sensitivity window of the FUV+24 \micron\ tracer \citep[e.g.,][]{kennicutt_star_1998, kennicutt_star_2012} and to remain within the regime where the SFHs are best constrained. We also
divide by the deprojected area to obtain a CMD-based SFR surface density, $\Sigma{\rm SFR}_{\rm CMD,100}$ as described in Section~\ref{sec:methods}, giving a consistent de-projection to a face-on SFR intensity as also used in Equation~\ref{eq:fuv_24mic}.

\begin{figure*}[ht]
    \centering
    \includegraphics[width=0.97\textwidth]{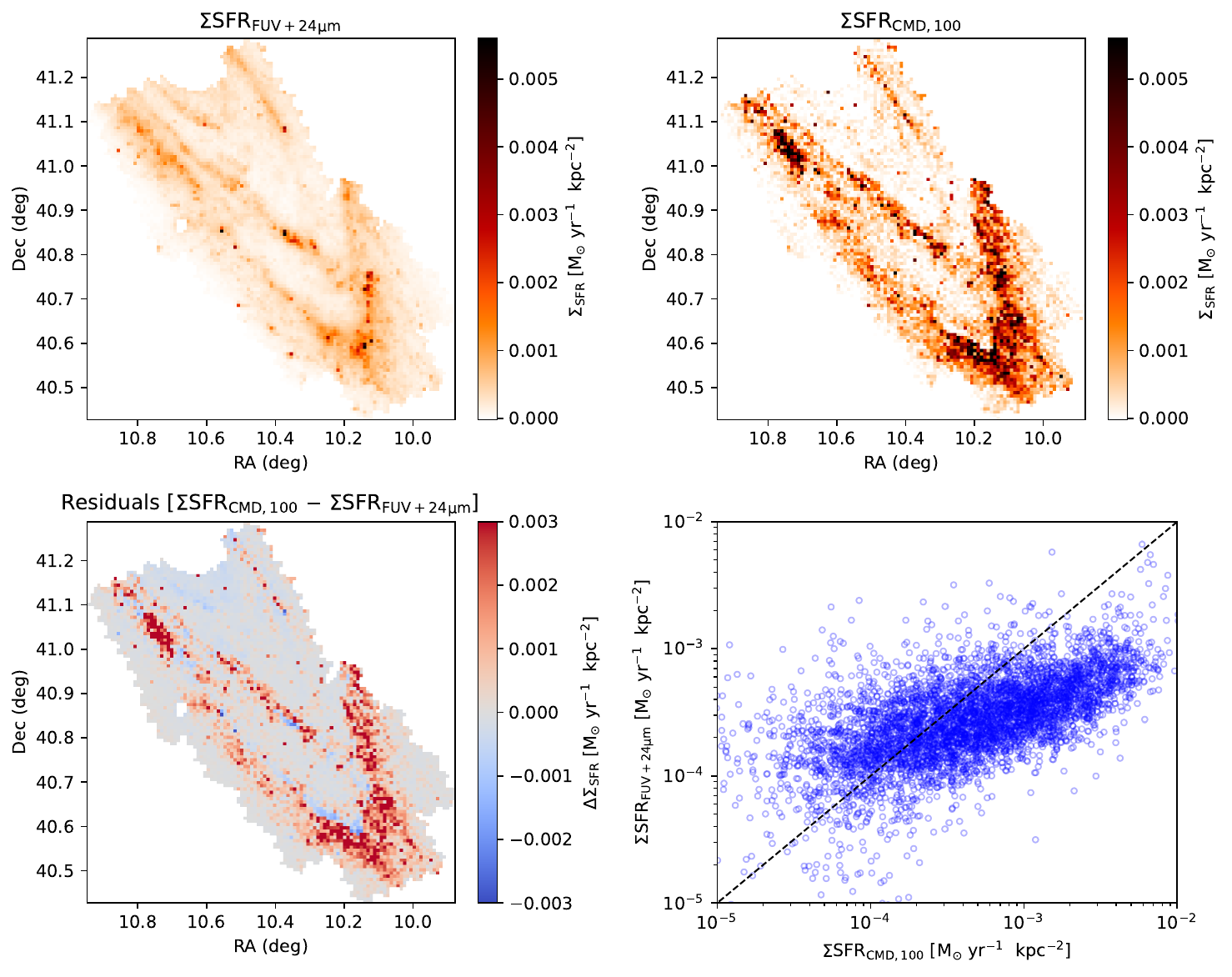}
    \caption{
    Comparison between SFR surface densities inferred from FUV+24 $\micron$ emission and from the CMD-derived SFHs. 
    Top left: SFR surface density map, $\Sigma{\rm SFR}_{\rm FUV+24}$, computed using the \citet{2008AJ....136.2782L} calibration applied to the \emph{GALEX} FUV and \emph{Spitzer}/MIPS 24 $\micron$ data \citep{martin_galaxy_2005, gordon_spitzer_2006}. 
    Top right: mean SFR surface density over the last 100~Myr from the CMD-based SFHs, $\Sigma{\rm SFR}_{\rm CMD,100}$, evaluated on the same spatial grid. 
    Bottom left: the difference between the two metrics, $\Sigma{\rm SFR}_{\rm CMD,100} - \Sigma{\rm SFR}_{\rm FUV+24}$, highlighting where the FUV+24 \micron\ predicts higher or lower SFRs than the CMDs. 
    Bottom right: pixel-by-pixel comparison of $\Sigma{\rm SFR}_{\rm FUV+24}$ versus $\Sigma{\rm SFR}_{\rm CMD,100}$, with the dashed line indicating equality. 
    The FUV+24 \micron\ systematically yields higher SFR surface densities, by factors of an order $\sim2.65$, while preserving the same large-scale morphology seen in the CMD-based map.
    }
    \label{fig:fuv24_vs_cmd}
\end{figure*}

We compare the SFRs derived from these two methods in Figure~\ref{fig:fuv24_vs_cmd}. The top-left panel shows the $\Sigma{\rm SFR}_{\rm{FUV}+24 \mu m}$ map, and the top-right panel shows $\Sigma{\rm SFR}_{\rm CMD,100}$ over the same footprint and at the same spatial resolution. The bottom-left panel shows the residuals,
and the bottom-right panel plots $\Sigma{\rm SFR}_{\rm CMD,100}$ versus $\Sigma{\rm SFR}_{\rm{FUV}+24 \mu m}$ for all spatial regions. 

Comparing the top two panels, the overall morphologies are remarkably similar. The 10~kpc ring stands out prominently in both maps, as do the star forming region NGC~206 and many of the localized peaks associated with bright OB associations. 

While the morphology of the SFR maps are similar between both methods, there is an obvious difference in scale. We find that the FUV+24 \micron\ SFR estimator yields systematically lower SFR surface densities than the CMD based SFRs. Specifically, we find that a multiplicative offset of 2.1 minimized the median absolute deviation (MAD) of squared residual (i.e., ($\Sigma{\rm SFR}_{\rm CMD,100}$ $-$ $\Sigma{\rm SFR}_{\rm{FUV}+24 \mu m}$)$^2$ )) across the map as a whole. 

Examining the residual map in the lower left panel of Figure~\ref{fig:fuv24_vs_cmd}, the bias towards lower FUV+24 \micron\ SFRs is clearest in the highest SFR regions in the ring structures. When we plot the direct region-by-region comparisons (bottom right panel), it is clear that the slope of the residuals does not follow the 1:1 relation. In fact, in log-log space as shown, a line of best fit has a slope of 0.34, where 1.0 would represent the 1:1 relation. Additionally, there is a significant scatter in the fractional residuals (defined to be [ ($\Sigma{\rm SFR}_{\rm{FUV}+24 \mu m}$ - $\Sigma{\rm SFR}_{\rm CMD,100}$) / $\Sigma{\rm SFR}_{\rm CMD,100}$ ] where the interquartile range (IQR) is 0.21.

This structure leads to larger offsets for more intense star formation (large $\Sigma{\rm SFR}_{\rm CMD,100}$). Similarly, at the very lowest SFRs, there is a slight tendency for the FUV+24 \micron\ SFRs to be higher than those inferred from the CMDs, although with significant scatter. Fundamentally, the dynamic range of the CMD-based SFRs appears to be far greater than those derived from the FUV+24 \micron\ data, with the former spanning an additional order of magnitude in $\Sigma_{\rm SFR}$. We explore potential origins of this offset below.

\subsection{The Only Way to Reproduce the Observed FUV Emission is With a Time Varying SFH} \label{sec:only_way}

The offsets shown in Figure~\ref{fig:fuv24_vs_cmd} reveal significant, systematic differences between the the CMD based 100 Myr average SFR, and SFRs derived from FUV+24 \micron\ emission. This pattern may indicate a mismatch in how the embedded, youngest star formation is treated in the two approaches, or, in the assumed timescales over which the SFR is being measured. We explore each of these possibilities in Appendix~\ref{appendix:IR_test}, and conclude that a possible culprit for the offset appears to be the timescale sensitivity of the FUV in a rapidly evolving recent SFH.

However, in Section~\ref{sec:synthetic_fuv}, we showed that the CMD-derived SFHs can reproduce the observed \emph{GALEX} FUV map across the PHAST footprint. In principle, that agreement might be reproduced by either of the constant SFRs presented in Section~\ref{sec:sfrsd_fuv24}, even if the full time variability of the SFH is ignored. Here we test that possibility directly by comparing synthetic FUV images constructed from constant recent SFRs (Section~\ref{sec:sfrsd_fuv24}) to the full time-resolved CMD-derived SFHs (Section~\ref{sec:synthetic_fuv}). This comparison then also tests whether the FUV emission is governed by the average recent star formation, or the timing of that star formation over the last 100 Myr.

We perform this test by repeating the \texttt{FSPS} based synthetic image generation described in Appendix~\ref{appendix_fsps}, but replace the last 100 Myr in each region with a constant SFR. We consider two choices for this constant value. In the first case, we adopt the CMD-based 100 Myr average, SFR$_{\rm CMD,100}$. In the second, we adopt SFR$_{\rm{FUV}+24 \mu m}$ from Figure~\ref{fig:fuv24_vs_cmd}. For both instances, we treat dust extinction the same as we did in Section~\ref{sec:synthetic_fuv}.

\begin{figure*}
    \centering
    \includegraphics[width=0.95\linewidth]{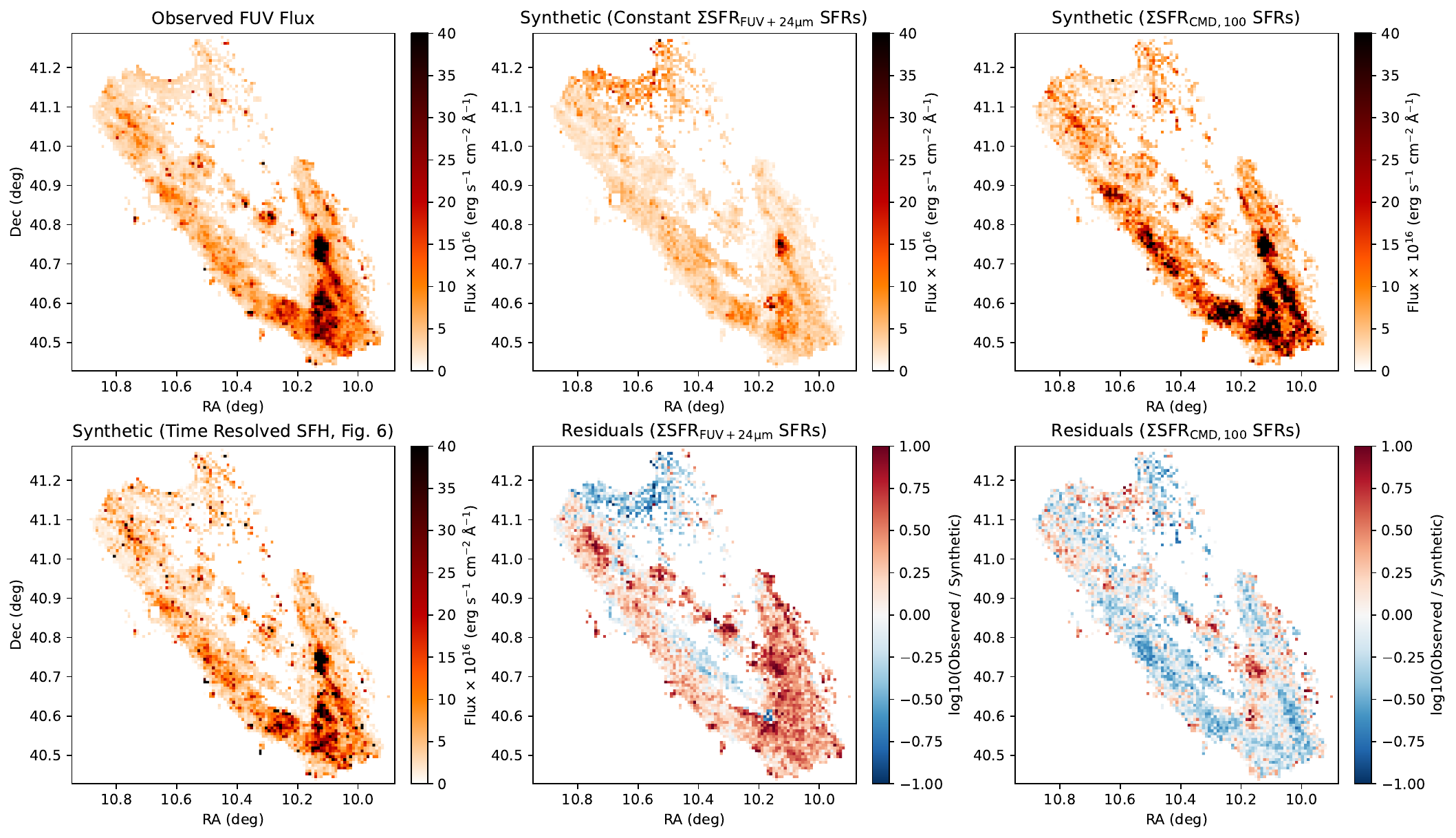}
    \caption{Comparison between the observed FUV emission and synthetic FUV maps constructed under different assumptions for the recent SFH. \textit{Top left:} observed \emph{GALEX} FUV flux, background-subtracted and measured on the PHAST spatial grid. \textit{Top middle:} synthetic FUV map generated by setting the SFR in the last 100 Myr in each region with a constant SFR equal to SFR$_{\rm{FUV}+24 \mu m}$. \textit{Top right:} synthetic FUV map generated by by setting the SFR in the last 100 Myr in each region with a constant SFR equal to the 0–100 Myr CMD-based average, SFR$_{\rm CMD,100}$. \textit{Bottom left:} synthetic FUV map generated from the full time-resolved CMD-derived SFHs, reproduced from Figure~\ref{fig:synthetic_fuv} for direct comparison. \textit{Bottom middle:} residual map for the constant-SFR$_{\rm{FUV}+24 \mu m}$ case, shown as $\log_{10}({\rm observed}/{\rm synthetic})$. \textit{Bottom right:} residual map for the constant-SFR$_{\rm CMD,100}$ case, shown in the same way. Both constant-SFR prescriptions fail to reproduce the observed FUV distribution as well as the full time-resolved SFHs.}
    \label{fig:constant_sux}
\end{figure*}

We present the results from these tests in Figure~\ref{fig:constant_sux}. The top-left panel shows the observed \emph{GALEX} FUV map measured on the PHAST spatial grid, while the top-middle and top-right panels show synthetic FUV images constructed after replacing the recent SFH in each region with a constant SFR equal to SFR$_{\rm{FUV}+24 \mu m}$ and the 0–100 Myr CMD-based average, respectively. For reference, the bottom-left panel reproduces the synthetic FUV map generated from the full time-resolved CMD-derived SFHs shown in Figure~\ref{fig:synthetic_fuv}. The two residual panels then show how the constant-SFR experiments depart from the observations.

These results demonstrate that a time-varying SFH is required to reproduce the observed FUV flux. In both of these tests, we fail to recover a synthetic FUV flux that is consistent with observations. Using a constant SFR set from the SFR$_{\rm{FUV}+24 \mu m}$ values leads to a MAD of the squared residuals of 4.71 from the true observed FUV map shown in the top left panel of Figure~\ref{fig:constant_sux} (and in Figure~\ref{fig:synthetic_fuv}). If the constant SFR is instead set to the (higher) average value from the CMD-based SFR over the past 100 Myr, the MAD of the squared residuals are 4.28, which is smaller, but only slightly. Both values are around $2.5\times$ worse than the residuals obtained when the full time-varying SFH is used (Figure~\ref{fig:synthetic_fuv}).  

Although both constant-SFR experiments perform poorly, they fail in systematically different ways. Setting the constant rate at the CMD-based 100 Myr average provides too much FUV light because more of the star formation during the interval has been pushed to more recent times, where the younger stellar ages will have larger contributions to the FUV for a given stellar mass. In this case, the fractional residual distribution [ (synthetic - observed) / observed ] is centered at 0.24, indicating that the synthetic flux is typically too high. 

In contrast, setting the constant rate to that inferred from the FUV+ 24 \micron\ method  (i.e. SFR$_{\rm{FUV}+24 \mu m}$ ) provides too little FUV flux compared to what is observed. In this case, the synthetic image systematically under-predicts the observed FUV emission, with the fraction residual distribution centered at -0.38.  The amplitude of this underestimate is roughly consistent with the factor of 2 under-prediction of the CMD-based SFR established earlier in Figure~\ref{fig:fuv24_vs_cmd} (see Section~\ref{sec:sfrsd_fuv24}). 

Overall, the failure of either constant SFR model to explain the observed FUV emission is due to the fact that the FUV is sensitive not just to the averaged rate of star formation over some time interval, but also depends on the detailed timing of star formation within that interval. Physically, the FUV emission is weighted most strongly toward short lived O and B stars with lifetimes shorter than 100 Myr, with a strong weighting toward ages of 30–50 Myr \citep[e.g.,][]{kennicutt_star_1998, kennicutt_star_2012, oti-floranes_calibration_2010}. This makes the FUV especially sensitive to the star formation rate over the last few to few tens of Myr, when the mass-to-light ratio is lowest. In contrast, when the star formation is primarily biased to later times, it takes more star formation to drive the observed flux, due to the higher mass-to-light ratio.

In the specific case of M31, the CMD-based SFR$_{\rm CMD,100}$ averaged over 0–100 Myr is higher than the actual SFR at any point during the last $\sim 40$\, Myr (Section~\ref{sec:full_sfh}). When a larger 100 Myr average is imposed as a constant recent SFR, the synthetic population contains too many young massive stars and therefore over-predicts the observed FUV flux. 

The FUV+24 \micron\ SFR constant value moves in the opposite direction. Because the FUV+24 \micron\ SFR indicator assumes a constant SFR when setting the conversion factor from FUV flux to an un-obscured SFR, it naturally assumes the presence of a significant fraction of young, low mass-to-light ratio stellar populations. However, in M31, these populations are not actually prevalent, because of the rapidly declining SFR in the galaxy. Because the FUV+24 \micron\ SFR values are systematically lower than the CMD-based 100 Myr average, imposing them as a constant recent SFR produces too little recent star formation and therefore too little FUV emission. In that case the deficit is not limited only to the most massive O and B stars, but also reflects an underproduction of the broader young stellar population contributing to the UV over the full $\sim 100$ Myr interval.

In summary, the observed FUV flux can only be reproduced from the CMD-derived SFHs in their full time resolution. These tests demonstrate that the observed FUV emission is sensitive not just to the amount of recent star formation, but to its detailed timing. 
Standard FUV-based calibrations implicitly assume a much steadier recent SFH than what our analysis recovers for M31, where the SFR has declined substantially over the last $\sim 100$ Myr. M31 therefore provides a clear example of how time variability in the recent SFH can produce a substantial systematic offset between CMD-based and FUV-based SFR estimates.

However, this declining SFR is unlikely to be the only contributor to the observed offset. Other studies have found non-negligible discrepancies, at the level of at least $\sim 50\%$ in predicted FUV flux, even in systems where the recent SFH appears comparatively steady over the last 100 Myr \citep{mcquinn_calibrating_2015,lazzarini_panchromatic_2022}. Additional effects, including uncertainties in stellar evolution models (Section~\ref{sec:isochrone_uncertainties}), IMF assumptions, photometric completeness (Section~\ref{sec:asts}), dust geometry and covering fraction, and assumptions about the fraction of UV light that is reprocessed into observable infrared emission, also contribute at a non-negligible level. These systematics imply that the shape of the recent SFH alone cannot explain the CMD verse FUV offset in every system. Our M31 results do however, clearly show that variations in the recent SFH can be a dominant source of systematic error when applying standard FUV-based SFR calibrations. 

The tension between the CMD-based and FUV+24 \micron\ SFR calibrators in M31 therefore appears to arise from the timescale over which FUV emission traces a time-varying recent SFH. This result underscores that absolute SFR calibrations in nearby galaxies still carry systematic uncertainties at the level of factors of a few, and more investigation will be needed on modeling the SFR across systems where systematic differences can be ruled out. For applications of the SFR in M31, we encourage readers to consider the full time varying SFH, opposed to a single, time averaged value.

\section{Summary} \label{sec: Summary}

In this paper we have used HST photometry from the PHAST survey \citep{chen_phast_2025} to map the recent star formation history across the southern disk of M31 at $\sim 100$~pc resolution. By fitting optical color-magnitude diagrams in more than 6500 regions, we recover spatially resolved SFHs over the past $\sim 500$~Myr that reveal the ring and arm structure of the southern disk and quantify how the recent SFR varies in both space and time.

Because the PHAST analysis follows the same \texttt{MATCH}-based methodology as the northern PHAT survey, we can combine the two data sets into a single, internally consistent view of recent star formation over roughly two thirds of M31's star-forming disk. 
We summarize our main results as follows:

\begin{itemize}
    \item \textit{PHAST SFR:}
    Across the PHAST footprint, which covers a deprojected area of $\sim 2.7\times10^2$~kpc$^2$ in the southern disk, the mean SFR over the last 100~Myr is $0.213 \pm 0.004~{\rm M_\odot~yr^{-1}}$, with most recent star formation concentrated in the 10 kpc ring, the inner arm structure, and NGC~206.

    \item \textit{A unified view over two thirds of the M31 disk from resolved stars}: Within the joint PHAT+PHAST footprint, which covers roughly two thirds of M31’s star-forming disk, we measure a mean SFR of $0.445 \pm 0.006~{\rm M_\odot~yr^{-1}}$ over the last 100~Myr. Extrapolating this value to the full star-forming disk, implies a SFR of $\sim 0.67~{\rm M_\odot~yr^{-1}}$ for M31’s total SFR.

    \item \textit{M31's global decline in star formation:}
    Both the PHAT and PHAST maps show that the recent SFH is not constant in time. The CMD-based SFHs indicate a clear, galaxy-wide decline in the mean SFR from $~1 {\rm M_\odot~yr^{-1}}$ $\sim 500$~Myr ago, decreasing to an average of $0.285 {\rm M_\odot~yr^{-1}}$ in the last 20 Myr. This decline appears to be the late stages of a multi Gyr decline in star formation from a more active state. Because the rings host the majority of the recent star formation, changes in their activity have a predominant impact on the disk-averaged SFH, and account for a substantial fraction of the observed decline.

    \item \textit{Quenching of the M32 region relative to the global average:}
    Within the PHAST footprint, the region surrounding M32 shows a distinct recent SFH compared to the disk as a whole. From 500~Myr to $\sim 60$~Myr ago, the M32 region is statistically consistent with the global average SFR. At lookback times younger than $\sim 60$~Myr, however, the SFR in this region drops well below the disk average, reaching a suppression of roughly a factor of three around 10–15~Myr ago. This localized quenching, adjacent to regions that show enhanced recent activity (such as the inner arm segment), highlights the strong environmental variations in M31’s recent SFH on kiloparsec scales, possibly related to interacting with M32.

    \item \textit{Comparison to FUV+24 \micron\ tracers:}
    We compare the CMD-based SFR surface densities, averaged over the last 100~Myr, to those inferred from FUV+24 \micron\ calibrations. Consistent with previous work, we find that the FUV based tracer systematically underestimates the CMD-based SFRs by a factor of approximately 2.1. At the same time, when we forward-model the PHAST SFHs and extinction parameters into a synthetic \emph{GALEX} FUV image, the resulting map reproduces the observed FUV morphology quite well, which cannot be done with a constant SFR. This agreement suggests that the CMD-derived SFHs provide a self-consistent description of where and when recent star formation has occurred. The mismatch with FUV+24 \micron\ estimates underscores that tracers implicitly averaged over $\sim100$ Myr are not reliable when the recent SFR is evolving over that time.
\end{itemize}

These results provide the most complete resolved-star view to date of recent star formation across M31, combining PHAT and PHAST into a nearly contiguous map over two thirds of the star-forming disk.

\vspace{5mm}


\begin{acknowledgments}
Support for this work was provided by NASA through grant \#GO-16778, 16796-16801 from the Space Telescope Science Institute, which is operated by the Association of Universities for Research in Astronomy, Incorporated, under NASA contract NAS5-26555. 
\end{acknowledgments}

\facilities{HST(ACS, WFC3)}

\software{\texttt{astropy} \citep{astropy_collaboration_astropy_2013, astropy_collaboration_astropy_2018, astropy_collaboration_astropy_2022}, \texttt{Jupyter} \citep{kluyver2016jupyter}, \texttt{matplotlib} \citep{Hunter:2007}, \texttt{numpy} \citep{harris_array_2020}, \texttt{python} \citep{python}, \texttt{scipy} \citep{2020SciPy-NMeth, scipy_12522488}, \texttt{h5py} \citep{collette_python_hdf5_2014, h5py_7560547}, \texttt{MATCH} \citep{dolphin_numerical_2002}, \texttt{FSPS} \citep{conroy_fsps_2010}
\citep{openai2023gpt4}}


\bibliographystyle{aasjournal}
\bibliography{references.bib, kam_agn_refs, software}

\appendix

\section{Creating Synthetic FUV images with \texttt{FSPS}}\label{appendix_fsps}

In Sections~\ref{sec:synthetic_fuv}, and \ref{sec:only_way}, we use the population synthesis code \texttt{FSPS} to generate synthetic FUV images. Here we describe the procedure used to derive these images. 

We feed the MATCH-derived SFHs directly into \texttt{FSPS} using its tabular SFH mode. The MATCH SFHs are specified as lookback-time bins, each with constant SFR. We convert these bins to an absolute-time grid, re-normalizing the youngest bin so that it extends to zero lookback while conserving the total stellar mass formed, and treat the SFR as piecewise constant in each interval. For each region we adopt a single metallicity equal to the SFR-weighted mean [M/H] over the last 100 Myr, or the metallicity of the most recent non-zero SFR bin, and assume this value for the full SFH. We then evaluate the \texttt{FSPS} model at $t_{\rm age} = 13.8$ Gyr to obtain integrated, dust-free \emph{GALEX} FUV and NUV AB magnitudes. We note that in following \citet{lewis_panchromatic_2017}, we use the full time resolution of our SFH fits out to log (age / yr) of 10.15, which is required to accurately model the UV flux given that $\sim20\%-30\%$ of the FUV emission comes from stars that are older than our adopted 500 Myr reliability threshold. A full discussion of this decision can be found in Appendix A of \citet{lewis_panchromatic_2017}.

Dust attenuation is modeled using the Milky Way–like \citet{cardelli_relationship_1989} law with R$_V$ = 3.1. To mimic the two-parameter extinction model used in the CMD fitting \citep[][ Section 2.2]{lewis_panchromatic_2015}, we draw a set of 20 extinction values uniformly between $A_V$ and $A_V$ + $dA_V$ (from Section~\ref{fig:av_dav}), recompute the FUV and NUV magnitudes for each draw, convert each realization to $f_\nu$, and average in flux space. The resulting intrinsic and attenuated magnitudes are then converted to fluxes using the \emph{GALEX} zero-points and placed on the PHAST spatial grid to construct synthetic UV images.

For the observed \emph{GALEX} image, we subtract a uniform background/sky as described above. Before comparing the observed image to the simulated one, however, we choose to restrict our comparison only to well-measured image pixels, taken to be those that are more than 3$\times$ the background noise in the real \emph{GALEX} image, when integrated over the area covered by the overlapping PHAST analysis region. We estimate an overall per–analysis-region $1\sigma$ uncertainty from the global background root mean square, scaled by the effective number of observed \emph{GALEX} image pixels covering the PHAST analysis region. We then compare this to the flux in the background–subtracted \emph{GALEX} FUV image integrated over the matching simulated image pixel in the simulated PHAST image.
We then only retained regions where the observed FUV flux exceeded the SNR=3 threshold.

\section{Pushing to Older Ages}\label{appendix}

Throughout this work, we have considered 500 Myr to be the limiting age where we can reliably recover the main sequence turnoff in our CMDs  (Section~\ref{sec:age_range_and_reliability}). This threshold is the same limit used in \citetalias{lewis_panchromatic_2015}, allowing for a seamless comparison between the analysis in the North and South. Additionally, this conservative limit allows us to maximize the amount of the disk we can reliably measure, and allows us to fully model the interior disk of M31 to look for Radial trends (Figure~\ref{fig:radial}). 

However, as we can see in Figure~\ref{fig:completeness_vs_density}, $\sim89\%$ of the PHAST survey has a density lower than 1.5 stars per arcsecond$^2$. From Figure~\ref{fig:turnoff}, we can see that in these less dense regions, we can therefore reliably probe to somewhat older ages than what is considered in the main body of the manuscript. 

\begin{figure}[ht]
    \centering
    \includegraphics[width=0.48\textwidth]{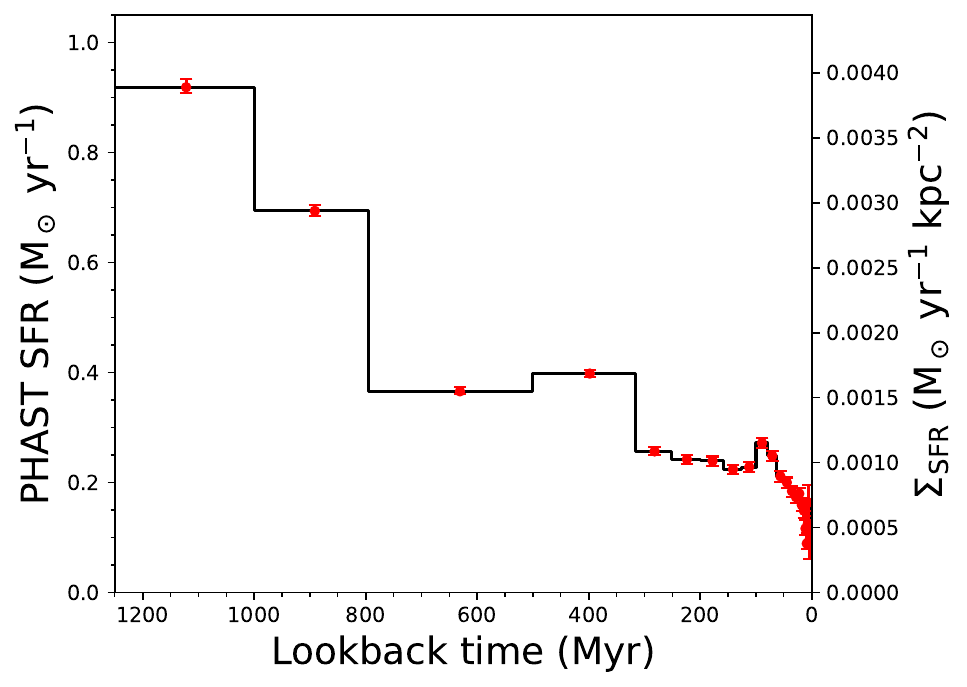}
    \caption{The SFH for PHAST regions with a density less than 1.5 stars per arcsecond$^2$, where our completeness limits allow us to probed older ages. Here we extend back to 1.2 Gyr, where we can reliably detect the MS turnoff above our completeness limit. We see a long decline in the SFR over this period of time.  }
    \label{fig:older}
\end{figure}

Shown in Figure\ref{fig:older}, are the SFH results for all PHAST regions with density lower than 1.5 stars per arcsecond$^2$, extending back to 1.2 Gyr.  These results are in striking agreement with the ancient SFH seen in PHAT \citep{williams_phat_2017}. Additionally, these results help anchor Section~\ref{sec: global_decline}, where we are observing a decline over several hundred Myr, extending back to the burst in SF M31 experienced 2-4 Gyr ago. What we observe in the last 500 Myr should then be interpreted as the tail of a longer decline.

\section{Testing Whether the Offset with FUV+ 24 \micron\ Tracers is Due in Part to the Handling of the IR and Dust Reprocessing} \label{appendix:IR_test}

A natural explanation for the observed offset between the CMD derived 100 Myr average SFR, and FUV+ 24 \micron\ derived SFRs seen in Section~\ref{sec:other_tracers} is that it arises in the infrared 24 \micron\ correction term, which is the component least directly constrained by the stellar populations measured in the CMD. The FUV+24 \micron\ calibration assumes that the 24 \micron\ emission provides a reliable correction for the UV light absorbed and reprocessed by dust. However, this assumption can break down when the SFR is changing. In particular, 24 \micron\ emission is dominated by dust heated by very young, deeply embedded massive stars and responds on timescales of only a few Myr \citep[e.g.,][]{kruijssen_uncertainty_2018, kim_duration_2021,wainer_timescales_2026}, whereas the CMD-derived SFRs are averaged over the last 100 Myr. In a system with a declining recent SFH, these two quantities are therefore not expected to match. In M31, where the SFR has declined significantly over the last $\sim 40$ Myr, the IR-based correction term should naturally contribute less than what would be expected for a constant-SFR population, and be systematically lower than the 100 Myr-averaged SFR inferred from the resolved stellar populations, even if both methods are internally consistent. It is therefore possible that the FUV+24 \micron\ SFRs can become systematically offset from the CMD-based SFHs even when both are tied to the same recent stellar populations. Here we test this hypothesis directly.

First, we consider whether a simple recalibration of the FUV+24 \micron\ tracer could reconcile it with the CMD-based SFRs, we explored modifying the relative wavelength components in the \citet{2008AJ....136.2782L} formula in Equation~\ref{eq:fuv_24mic}. Specifically, we re-fit the coefficients in Equation~\ref{eq:fuv_24mic} to produce the CMD-based SFRs, while retaining its functional form. 
Equation~\ref{eq:fuv_24mic} is then taken to be,
\begin{equation}
\Sigma{\rm SFR}_{\rm{FUV}+24 \mu m} = \Bigl( a_{\rm FUV}\, I_{\rm FUV}  + a_{24}\, I_{24} \Bigr) \cos i,
\end{equation}
and we re-fit the coefficients $a_{\rm FUV}$ and $a_{24}$.

We find best-fit values of $a_{\rm FUV}=0.36$ and $a_{24}=3.1\times10^{-3}$. This refit substantially reduces the global normalization offset and improves the slope of the region-to-region residual, where the best-fit slope shifts to 0.42, closer to the 1.0 expected for the one-to-one relation, but still dramatically off base. Additionally, the improvement is mostly in the overall trend rather than in the local agreement. The MAD of the squared residuals is reduced by only $\sim 3\%$ relative to the original \citet{2008AJ....136.2782L} calibration, and the fractional residual distribution remains similarly broad, with an IQR of 0.165.

Therefore, refitting the coefficients improves the offset without addressing the underlying disagreement. In practice, the new coefficients bring the highest-$\Sigma_{\rm SFR}$ regions into better agreement at the expense of over-predicting more of the lower-$\Sigma_{\rm SFR}$ regions, and the remaining discrepancies appear as spatially coherent departures across the disk. This behavior argues against the offset being explained by a single missing scale factor or a simple renormalization of the FUV and 24 \micron\ terms. Instead, it suggests that the CMD-based and FUV based estimates are responding differently to the recent, time-variable SFH and to the local dust geometry.

Interestingly, the main change in the refit occurs in $a_{\rm FUV}$, which increases by roughly a factor of four, while the best-fit $a_{24}$ remains broadly consistent with the original calibration. This results points to the FUV term, rather than the 24 \micron\ term, as the dominant contributor to the mismatch, a point we return to in Section~\ref{sec:only_way}. In fact, the regions with high levels of 24 \micron\ emission are not the same as those with FUV emission, and any adjustment to the $a_{24}$ terms results in notably worse scatter in the residuals. 


Another possibility is that the problem lies specifically with the 24 \micron\ band itself. If that were the case, one might try replacing the 24 \micron\ term in Equation~\ref{eq:fuv_24mic} with a longer-wavelength IR band or with a total-IR estimate \citep[e.g.,][]{boquien_towards_2016}. However, a growing body of work suggests that both mid–IR and especially far–IR emission can be substantially powered by dust heated from evolved stellar populations rather than by the youngest massive stars \citep[e.g.,][]{li_spitzer_2010,bendo_investigations_2012,groves_heating_2012, whitcomb_star_2023}. In nearby spirals, analyses of the radial energy budget find that a significant fraction of the 24 \micron\ flux, particularly in bulges and inner disks, arises from dust heated by older stars rather than embedded H\textsc{ii} regions \citep[e.g.,][]{boquien_dust_2011,leroy_estimating_2012}. At longer wavelengths, \emph{Herschel} studies show that the bulk of the emission beyond $\sim 160,\micron$ is often dominated by heating from the diffuse old stellar population, with only a minority of the far–IR luminosity directly linked to current star formation \citep[e.g.,][]{smith_herschel_2012,viaene_herschel_2017}.
\footnote{In M31 specifically, radiative–transfer modeling demonstrates that evolved stars dominate the heating of dust over most of the disk at $\lambda \gtrsim 100,\micron$, with young stars becoming energetically important only in the star–forming ring \citep[e.g.,][]{viaene_herschel_2017}.}

These results imply that simply switching from 24 \micron\ to longer–wavelength IR, or to a total–IR term, would introduce an even larger and more spatially variable contribution from dust heated by old stars, rather than providing a clean fix to the discrepancy with CMD–based SFRs. Any attempt to recalibrate the FUV based indicators therefore has to explicitly account for both the timescale sensitivity of the FUV+IR combination and the local balance between young– and old–star dust heating, rather than assuming a single universal IR correction.

The tests shown here suggest that the offset observed between $\Sigma{\rm SFR}_{\rm CMD,100}$ and $\Sigma{\rm SFR}_{\rm{FUV}+24 \mu m}$ are not primarily caused by an incorrect treatment of embedded star formation traced by in the IR by dust-reprocessing. Although re-scaling the terms in Equation~\ref{eq:fuv_24mic} can reduce the global normalization offset, it does not remove the spatially coherent residuals or substantially reduce the scatter. The primary culprit therefore appears to be the differing timescale sensitivity of the FUV in a rapidly evolving recent SFH.

\end{document}